\documentclass[aps,superscriptaddress,12pt,reprint,floatfix,amsmath]{revtex4-1}

\usepackage{graphicx}		
\usepackage[version=3]{mhchem}

\usepackage{color}                                                              %
\usepackage[scaled=.92]{helvet}                                                 %
\usepackage{caption}                                                            %
\usepackage{sfmath}                                                             %
\setcitestyle{super}                                                            %
\makeatletter                                                                   %
\renewcommand\NAT@citesuper[3]{\ifNAT@swa                                       %
\unskip\kern\p@\textsuperscript{\NAT@@open #1\NAT@@close}%                      %
\if*#3*\else\ (#3)\fi\else #1\fi\endgroup}                                      %
\makeatother                                                                    %
\renewcommand{\onlinecite}[1]{\hspace{-1 ex} \nocite{#1}\citenum{#1}}           %
\begin{document}

% Use the \preprint command to place your local institutional report number
% on the title page in preprint mode.
% Multiple \preprint commands are allowed.
%\preprint{}

\title{A singlet and triplet excited-state dynamics study of the keto and enol tautomers of cytosine} %Title of paper

% repeat the \author .. \affiliation  etc. as needed
% \email, \thanks, \homepage, \altaffiliation all apply to the current author.
% Explanatory text should go in the []'s,
% actual e-mail address or url should go in the {}'s for \email and \homepage.
% Please use the appropriate macro for the type of information

% \affiliation command applies to all authors since the last \affiliation command.
% The \affiliation command should follow the other information.

\author{Sebastian Mai}
\affiliation{Institute of Theoretical Chemistry, University of Vienna, W\"ahringer Str. 17, 1090 Vienna, Austria}

\author{Philipp Marquetand}
\email[]{philipp.marquetand@univie.ac.at}
\affiliation{Institute of Theoretical Chemistry, University of Vienna, W\"ahringer Str. 17, 1090 Vienna, Austria}

\author{Martin Richter}
\affiliation{Institute of Theoretical Chemistry, University of Vienna, W\"ahringer Str. 17, 1090 Vienna, Austria}
\affiliation{Institute of Physical Chemistry, Friedrich-Schiller-Universit\"at Jena, Helmholtzweg 4, 07743 Jena, Germany}

\author{Jes\'us Gonz\'alez-V\'azquez}
\affiliation{Departamento de Qu\'{\i}mica, M\'odulo 13, Universidad Aut\'onoma de Madrid, Cantoblanco, 28049 Madrid, Spain}

\author{Leticia Gonz\'alez}
\affiliation{Institute of Theoretical Chemistry, University of Vienna, W\"ahringer Str. 17, 1090 Vienna, Austria}

% Collaboration name, if desired (requires use of superscriptaddress option in \documentclass).
% \noaffiliation is required (may also be used with the \author command).
%\collaboration{}
%\noaffiliation

\date{\today}

\begin{abstract}
The photoinduced excited-state dynamics of the \emph{keto} and \emph{enol} forms of cytosine is investigated using ab initio surface hopping in order to understand the outcome of molecular beam femtosecond pump-probe photoionization spectroscopy experiments. Both singlet and triplet states are included in the dynamics. The results show that triplet states play a significant role in the relaxation of the \emph{keto} tautomer, while they are less important in the \emph{enol} tautomer. In both forms, the $T_1$ state minimum is found too low in energy to be detected in standard photoionization spectroscopy experiments and therefore experimental decay times should arise from a simultaneous relaxation to the ground state and additional intersystem crossing followed by internal conversion to the $T_1$ state. In agreement with available experimental lifetimes, we observe three decay constants of 7 fs, 270 fs and 1900 fs -- the first two coming from the \emph{keto} tautomer and the longer one from the \emph{enol} tautomer. Deactivation of the \emph{enol} form is due to internal conversion to the ground state via two $S_1/S_0$ conical intersections of ethylenic type.
%The long excited-state lifetime of the \emph{enol} in comparison to the \emph{keto} tautomer is due to differences in the electronic structure induced by the absence of the carbonyl group.
\end{abstract}

% insert suggested PACS numbers in braces on next line

\maketitle %\maketitle must follow title, authors, abstract and \pacs

% =========================================================================================================================================== %
% =========================================================================================================================================== %
% =========================================================================================================================================== %

\section{Introduction}\label{sec:introduction}

Since the advent of ultrafast time-resolved spectroscopy and modern \emph{ab initio} methods, the electronic structure and excited-state femtosecond nuclear dynamics of the five nucleobases, which are the central building blocks of DNA and RNA, have been studied intensively.\cite{crespo-hernandez_ultrafast_2004}
Understanding these fundamental processes can help unraveling the mechanisms giving DNA and RNA a remarkable resistance against damage from ultraviolet (UV) irradiation. It is generally accepted that all nucleobases undergo an ultrafast (i.e. on a picosecond or shorter timescale) relaxation to the electronic ground state upon excitation by UV light, thereby spreading excess energy among the different degrees of freedom before any harmful reaction can occur.

\begin{figure}
  \centering
  \includegraphics[width=246pt]{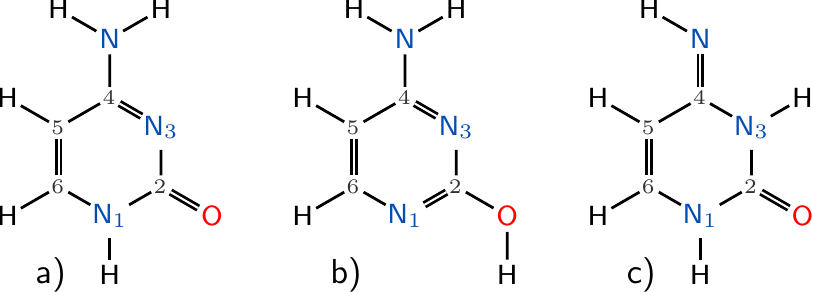}
  \caption{Tautomers of cytosine. In a) the \emph{keto} form, in b) the \emph{enol} form and in c) the \emph{imino} form.}
  \label{fig:tautomers}
\end{figure}

Among the nucleobases, the excited-state dynamics of cytosine has attracted considerable attention. Cytosine exists primarily in three tautomers, the \emph{keto}, \emph{enol} and \emph{imino}-forms, see Fig.~\ref{fig:tautomers}. Since the \emph{keto} tautomer is the biologically relevant one and the only one found in aqueous solution and solid state, most spectroscopic efforts focus on identifying the relaxation mechanism of this particular tautomer. However, since the \emph{enol} form is dominant in gas phase and the \emph{imino} form can also be present depending on the experimental conditions, the interpretation of studies dealing with the excited-state dynamics of \emph{keto} cytosine can get severely complicated.

Several experimental and theoretical studies have been aimed at identifying the tautomer ratios in gas phase. Brown et al.\cite{brown_tautomers_1989} reported a tautomer ratio of 0.44:0.44:0.12 (\emph{keto}:\emph{enol}:\emph{imino}), determined by microwave spectroscopy. Szczesniak et al.\cite{szczesniak_matrix_1988} detected a ratio of 0.32:0.65:0.03 in matrix-isolation infrared (IR) studies. More recently, Bazs\'o et al.\cite{bazso_tautomers_2011} measured the tautomer ratio from matrix-isolation IR and UV spectra, obtaining 0.22:0.70:0.08.

The excited-state lifetime of cytosine in gas phase has been measured in ultrafast molecular beam experiments with quite different outcomes depending on the particular experimental setup. Kang et al.\cite{kang_intrinsic_2002} (pump: 267 nm, probe: multiphoton 800 nm) observed a single exponentially decaying signal with a time constant of 3.2 ps. Canuel et al.\cite{canuel_excited_2005} (pump: 267 nm, probe: 2$\times$400 nm) identified two independent time constants of 160 fs and 1.86 ps. Ullrich et al.\cite{ullrich_electronic_2004} (pump: 250 nm, probe: 200 nm) even resolved three components, one extremely fast at 50 fs, one at 820 fs and a third one at 3.2 ps. More recently, Kosma et al.\cite{kosma_excited-state_2009} measured the excited-state lifetime using pump wavelengths between 260 and 290 nm (probe: 3$\times$800 nm) and showed that the excited-state lifetime strongly varies with the excitation energy.
For short wavelengths (below 280 nm), their results are in line with earlier findings: At 260 nm two transients are identified with an ultrafast decay of 120 fs followed by a slower relaxation path at 3.8 ps; with excitation at 267 nm and 270 nm, similar time scales are found but a third small longer transient is also observed. For wavelengths of 280 and 290 nm, the first transient is below 100 fs, the second is around 1 ps, and the third is very long-lived (55 and above 150 ps, respectively).
%For wavelengths of 290 and 280 nm, they found both extremely fast components with time constants below 100 fs as well as very long-lived components (above 100 ps). For short wavelengths, their results are more in line with the earlier findings, i.e. components with time constants of several 100 fs as well as a few ps.
Lately, Ho et al.\cite{ho_disentangling_2011} investigated the different excited-state lifetimes of the three main tautomers using derivate molecules. In 1-methylcytosine (not possessing the \emph{enol} form) they resolve time constants from 0.4 ps at 260 nm to 3.2 ps at 310 nm, while in 5-fluorocytosine (primarily in \emph{enol} form) they found one wavelength-independent component of 300 fs  along with another timescale of 9.5 to 100 ps, again depending on the pump wavelength. For cytosine itself, at 266 nm they resolved two time constants, one at 500 fs and another at 4.5 ps, while for longer wavelengths (290-300 nm) only one transient at ca 1 ps was found and the long-lived component vanished. In all systems they also detected an initial spike, which could hide a very fast dynamical process but could not be resolved. Kotur et al.\cite{kotur_following_2012,kotur_distinguishing_2011} measured excited state lifetimes of cytosine paying attention to separate different fragment ions and for the parent ion found three time constants at 50, 240 and 2360 fs, using a 262 nm excitation wavelength.

The most recent vertical excitation energies of cytosine tautomers have been reported by Tomi\'c et al.\cite{tomic_quantum_2005} (using DFT/MRCI, density functional theory/multireference configuration interaction), Blancafort~\cite{blancafort_energetics_2007} (CASSCF and CASPT2, complete active space self-consistent field and its second-order perturbation theory variant) and Szalay et al.\cite{tajti_reinterpretation_2009,bazso_tautomers_2011} (EOM-CCSD and CC3, equation-of-motion coupled-cluster including singles and doubles and the related coupled-cluster approach including triples) in gas phase, and in solution by Domingo and coworkers~\cite{domingo_absorption_2012} (CASPT2). Intensive theoretical work has been devoted to identify conical intersections (CoIns) in the singlet manifold of \emph{keto} cytosine. The recent studies of Kistler et al. located a number of two-state CoIns\cite{kistler_radiationless_2007} and three different three-state CoIns\cite{kistler_three-state_2008} using MRCI methods. Additionally, Barbatti et al.\cite{barbatti_photodynamical_2011} optimized four CoIns and a number of state minima, also at MRCI level. The earlier paper of Tomi\'c et al.\cite{tomic_quantum_2005} reports excited-state minima and CoIns at the DFT/MRCI level of theory. Based on quantum chemical results, several deactivation mechanisms for the \emph{keto} cytosine have been proposed by different authors.\cite{ismail_ultrafast_2002,merchan_ultrafast_2003,blancafort_key_2004,merchan_triplet-state_2005,blancafort_singlet_2005,gonzalez-luque_singlettriplet_2010,sobolewski_ab_2004,zgierski_origin_2005}

Only a limited number of dynamical simulations of the excited-state dynamics of the \emph{keto} tautomer has been conducted. Hudock and Mart\'{\i}nez\cite{hudock_excited-state_2008} used ab initio multiple spawning based on CASSCF(2,2), finding multiple subpisecond pathways involving only two electronic states. Lan and coworkers~\cite{lan_photoinduced_2009} used surface-hopping with the semiempirical OM2 method considering the two lowest excited singlet states. Gonz\'alez-V\'azquez and Gonz\'alez~\cite{gonzalez-vazquez_time-dependent_2010} and independently Barbatti et al.\cite{barbatti_photodynamical_2011} performed surface-hopping in an ab initio framework with four singlet states. The first study including triplet states in \emph{keto}-cytosine was reported by Richter et al.\cite{richter_femtosecond_2012} who showed that intersystem crossing (ISC) is taking place on an ultrafast time scale. To the best of our knowledge, no dynamics simulations concerning the \emph{enol} tautomer have been performed so far.

The present work is an attempt to provide new insights into the relaxation process of cytosine by studying the excited-state dynamics of both the \emph{keto} and the \emph{enol} tautomers. Regrettably, the \emph{imino} form could not be included, since preliminary studies suggested that the here employed level of theory for the \emph{keto} and \emph{enol} is not able to properly describe the excited states of the \emph{imino} tautomer. Additionally, the relative abundance of this tautomer is below 10\%, justifying the focus on the \emph{keto} and \emph{enol} forms. 
Both the \textit{enol} and \emph{imino} tautomers show two rotamers, depending on the orientation of the \ce{OH}- and the \ce{NH}-groups. For both tautomers, the lowest energy structure in gas phase\cite{bazso_tautomers_2011} is shown in Fig.~\ref{fig:tautomers}. Accordingly, the calculations on the \emph{enol} tautomer only included this rotamer.
Since triplet state formation has been proposed by several studies in \emph{keto} cytosine,\cite{nir_rempi_2002,merchan_triplet-state_2005,gonzalez-luque_singlettriplet_2010} the present study also includes the interaction between singlet and triplet states, using a methodology similar to the one employed in Ref.~\onlinecite{richter_femtosecond_2012}. As shown in Section II, the surface-hopping method presented here is more robust in the presence of weak spin-orbit couplings. Since its application leads to small differences  with respect to the results discussed in Ref.~\onlinecite{richter_femtosecond_2012}, the dynamics of the \emph{keto} form is revisited in this paper and compared to that of the \emph{enol} form.

% =========================================================================================================================================== %
% =========================================================================================================================================== %
% =========================================================================================================================================== %

\section{Methodology}\label{sec:methodology}

\subsection{Surface hopping including arbitrary couplings}\label{ssec:SHARC}

Surface-hopping dynamics as proposed by Tully~\cite{tully_molecular_1990} is usually carried out in a basis of the eigenfunctions of the molecular Coulomb Hamiltonian (MCH). Within the Born-Oppenheimer approximation, the corresponding electronic Hamiltonian contains the electronic kinetic energy and the potential energy arising from the Coulomb interaction of the electrons and nuclei with each other, i.e.,
\begin{equation}
  \hat{H}_{\text{el}}^{\text{MCH}}
  =\hat{K}_{\text{e}}
  +\hat{V}_{\text{ee}}
  +\hat{V}_{\text{ne}}
  +\hat{V}_{\text{nn}}.
\end{equation}
Standard quantum chemistry programs usually obtain wavefunctions as eigenfunctions of this operator, and a large number of properties can be calculated for these wavefunctions. However, the description of phenomena like light-matter interaction or ISC necessitate the use of additional terms in the Hamiltonian, e.g. dipole couplings or spin-orbit couplings:%
\begin{equation}
  \hat{H}_{\text{el}}^{\text{total}}
  =\hat{H}_{\text{el}}^{\text{MCH}}
  +\hat{H}_{\text{el}}^{\text{coup}}.
\end{equation}
Because of the classical approximations inherent to surface-hopping, integration of the nuclear motion should be performed on the potential energy surfaces (PESs) of the eigenfunctions of the total electronic Hamiltonian $\hat{H}_{\text{el}}^{\text{total}}$. However, as these eigenfunctions and their properties are usually not obtainable with quantum chemistry software, in the recently developed \textsc{Sharc} (Surface Hopping including ARbitrary Couplings) methodology~\cite{richter_sharc:_2011} we use the eigenfunctions of $\hat{H}_{\text{el}}^{\text{total}}$ in the subspace of the few lowest eigenstates of the MCH Hamiltonian. Henceforth, the basis of the eigenfunctions of $\hat{H}_{\text{el}}^{\text{total}}$ will be referred to as the diagonal basis, since this Hamiltonian is diagonal in this basis.

In order to obtain surface-hopping probabilities, the electronic wavefunction is expanded as a linear combination of the diagonal basis functions:%
\begin{equation}
  |\Psi_{\text{el}}\rangle=\sum_\alpha|\phi_\alpha^{\text{diag}}\rangle c_\alpha^{\text{diag}}.
  \label{eq:wavefunction}
\end{equation}
Inserting this wavefunction into the time-dependent Schr\"odinger equation leads to the differential equation governing the evolution of the coefficients:%
\begin{equation}
  \frac{\partial}{\partial t}\mathbf{c}^{\text{diag}}=
  -\left[
    \frac{\text{i}}{\hbar}\mathbf{H}^{\text{diag}}
    +\mathbf{K}^{\text{diag}}
  \right]
  \mathbf{c}^{\text{diag}},
  \label{eq:evolution}
\end{equation}
where the Hamiltonian matrix $\mathbf{H}^{\text{diag}}$ has elements $\langle\phi_\beta^{\text{diag}}|\hat{H}|\phi_\alpha^{\text{diag}}\rangle$ and the elements of matrix $\mathbf{K}^{\text{diag}}$ are the non-adiabatic couplings $\langle\phi_\beta^{\text{diag}}|\partial/\partial t|\phi_\alpha^{\text{diag}}\rangle$, which in this basis include  the transformed coupling $\hat{H}_{\text{el}}^{\text{coup}}$.

Since the propagation of the coefficients is not subject to the classical approximation, the solution of equation \eqref{eq:evolution} is independent of the representation  of $\mathbf{H}$ and $\mathbf{K}$ and thus equation~\eqref{eq:evolution} can instead be written as:%
\begin{equation}
  \frac{\partial}{\partial t}\mathbf{c}^{\text{diag}}=
  -\mathbf{U}^\dagger\left[
    \frac{\text{i}}{\hbar}\mathbf{H}^{\text{MCH}}
    +\mathbf{K}^{\text{MCH}}
  \right]\mathbf{U}
  \mathbf{c}^{\text{diag}},
  \label{eq:evolution2}
\end{equation}
where $\mathbf{U}$ is given by $\mathbf{U}^\dagger\mathbf{H}^{\text{MCH}}\mathbf{U}=\mathbf{H}^{\text{diag}}$.

In the current version of \textsc{Sharc}, equation \eqref{eq:evolution2} is integrated numerically for a small timestep $\Delta t$ by:
\begin{equation}
  \mathbf{c}^{\text{diag}}(t)=\underbrace{\mathbf{U}^\dagger(t)\text{e}^{-\left[\text{i}\mathbf{H}^{\text{MCH}}(t)/\hbar+\mathbf{K}^{\text{MCH}}(t)\right]\Delta t}\mathbf{U}(t_0)}_{\mathbf{A}(t_0,t)}\mathbf{c}^{\text{diag}}(t_0),
\end{equation}
where $\mathbf{A}(t_0,t)$ is the total propagator from time $t_0$ to time $t$. In this way, the transformation of small couplings $\hat{H}_{\text{el}}^{\text{coup}}$ into highly peaked non-adiabatic couplings is avoided, allowing for a much more stable propagation, compared to Ref.~\onlinecite{richter_sharc:_2011}. Note that the surface-hopping itself is still performed in the diagonal basis, which is the optimal representation for this step in the algorithm; see also  Ref.~\onlinecite{Granucci2012JCP}. The corresponding surface-hopping probabilities from the current classical state $\beta$ to another state $\alpha$ are then calculated according to:
\begin{multline}
  P_{\beta\rightarrow\alpha}=\left(1-\frac{|c_\beta^{\text{diag}}(t)|^2}{|c_\beta^{\text{diag}}(t_0)|^2}\right)\\
  \times\frac{\Re\left[c^{\text{diag}}_\alpha(t)A^*_{\alpha\beta}(c^{\text{diag}}_\beta)^*(t_0)\right]}{|c^{\text{diag}}_\beta(t_0)|^2-\Re\left[c^{\text{diag}}_\beta(t)A^*_{\beta\beta}(c^{\text{diag}}_\beta)^*(t_0)\right]}.
\end{multline}
This is a modification of the equation derived by Granucci et al.\cite{granucci_direct_2001} used in the Local Diabatization algorithm\cite{plasser_surface_2012} available in~\textsc{Newton-X}.\cite{NEWTON-X} We also include decoherence as proposed in Ref.~\onlinecite{granucci_critical_2007} to the diagonal states.

% =========================================================================================================================================== %

\subsection{Ab initio level of theory and dynamics}\label{ssec:abinitio}

For both tautomers, the ground state equilibrium geometry was optimized using MP2/6-311G**\cite{krishnan_selfconsistent_1980} and harmonic frequencies were obtained at the same level of theory. From the obtained frequencies, a quantum harmonic oscillator Wigner distribution\cite{schinke_photodissociation_1995,dahl_morse_1988} was calculated and 2000 (1000 for the \emph{enol} form) geometries were sampled from the distribution. An absorption spectra is simulated employing the SA10S-CASSCF(12,9)/6-31G* level of theory~\cite{gonzalez-vazquez_time-dependent_2010}, where SA10S indicates that the calculation is averaged over 10 singlet states for each of the generated geometries.

The most typical excitation wavelength in the available experiments is 267~nm (4.64~eV), corresponding to the maximum of the first absorption band of the cytosine UV spectrum. Therefore, the center of our excitation energy range was chosen to be 5.07 eV, which is the maximum of the simulated composite spectrum given below. The band width was fixed at $\pm$0.07 eV, which is the typical energy band width of a 50 fs laser pulse.

From the generated sets of geometries,  initial conditions for the dynamics were selected, based on the oscillator strengths and the excitation energy and according to Refs. \onlinecite{barbatti_--fly_2007,NEWTON-X}.
For the \emph{keto} tautomer, 68 initial conditions were selected, 30 starting in the $S_1$, 36 in the $S_2$ and 2 in $S_3$; these are the most important states in the chosen energy range.
For the \emph{enol} tautomer, 65 initial conditions were considered (57 in $S_1$, 8 in $S_2$).
Subsequently, all 133 initial conditions were propagated with \textsc{Sharc}. Energies, gradients, non-adiabatic couplings and spin-orbit couplings were calculated on-the-fly using  the CASSCF(12,9)/6-31G* level of theory. In the case of the \emph{enol} tautomer, 3 singlet and 4 triplet states were averaged in the on-the-fly CASSCF procedure (denoted as SA3S+4T-CASSCF(12,9)/6-31G*), while for the keto tautomer 4 singlet and 3 triplet states were included (denoted as SA4S+3T-CASSCF(12,9)/6-31G*). The dynamics were simulated for 1000 fs (timestep of 0.5 fs, integration timestep 0.02 fs) or until relaxation to $S_0$ or $T_1$ occurred. The simulations take each of the triplet state components separately into account (i.e. the simulations consider 13 states for the \emph{keto} form and 15 states for the \emph{enol} form).

Using the geometries where surface hops between two states occurred, optimizations of CoIns or singlet-triplet crossings in the \emph{keto} and \emph{enol} forms of cytosine were carried out using the SA4S+3T- and SA3S+4T-CASSCF(12,9)/6-31G* level of theory, respectively. All the quantum chemical calculations have been carried out with the quantum chemistry package \textsc{Molpro} 2012.1.\cite{MOLPRO-WIREs,MOLPRO,WK85,KW85} Orbital visualization was done with the \textsc{Molekel 5.4} software.\cite{molekel}

% =========================================================================================================================================== %
% =========================================================================================================================================== %
% =========================================================================================================================================== %

\section{Results and Discussion}\label{sec:results}

% =========================================================================================================================================== %

\subsection{Spectra}\label{ssec:spectrum}

\begin{table}
  \centering
  \caption{Excitation energies $E_{\text{exc}}$ (eV) and oscillator strength $f_{0\alpha}$ for the low-lying excited states of the \emph{keto} and \emph{enol} tautomers at the Franck-Condon point calculated with SA10S+10T-CASSCF(12,9)/6-31G* (averaged over 20 states: 10 singlets and 10 triplets).}
  \begin{tabular}{lccclcc}
    \hline
    State $\alpha$ &$E_{\text{exc}}$      &$f_{0\alpha}$    &Character      &State $\alpha$  &$E_{\text{exc}}$ &Character\\
    \hline
    \multicolumn{7}{c}{--- \emph{keto} ---}\\
    $S_1$       &5.13   &0.0805 &$\pi\pi^*$     &$T_1$       &3.64   &$\pi\pi^*$\\
    $S_2$       &5.26   &0.0018 &$n\pi^*$       &$T_2$       &4.88   &$\pi\pi^*$\\
    $S_3$       &5.59   &0.0035 &$n\pi^*$       &$T_3$       &5.07   &$n\pi^*$\\
    $S_4$       &6.30   &0.0004 &$n\pi^*$       &$T_4$       &5.40   &$n\pi^*$\\
    \hline
    \multicolumn{7}{c}{--- \emph{enol} ---}\\
    $S_1$       &5.19   &0.0444 &$\pi\pi^*$     &$T_1$       &4.28   &$\pi\pi^*$\\
    $S_2$       &5.66   &0.0146 &$n\pi^*$       &$T_2$       &4.95   &$\pi\pi^*$\\
    $S_3$       &6.57   &0.0004 &$n\pi^*$       &$T_3$       &5.30   &$n\pi^*$\\
    &&&                                         &$T_4$       &5.43   &$\pi\pi^*$\\
    &&&                                         &$T_5$       &6.26   &$n\pi^*$\\
    \hline
  \end{tabular}
  \label{tab:exc}
\end{table}

\begin{figure}
  \centering
  \includegraphics[width=246pt]{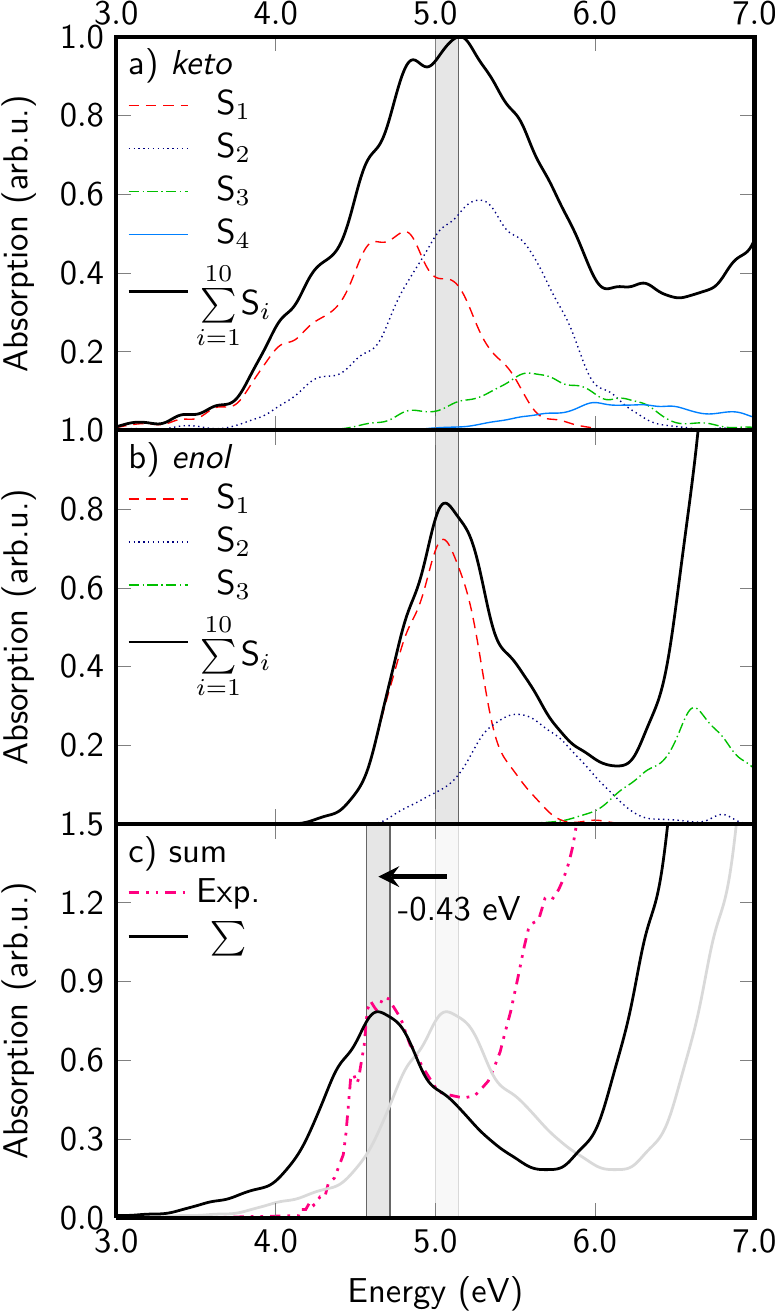}
  \caption{Simulated absorption spectra of the \emph{keto} (a), \emph{enol} (b) tautomers and a mixture of them (c). The composite spectrum (c), calculated as $0.76\cdot\sigma_{\text{enol}}+0.24\cdot\sigma_{\text{keto}}$ and shifted by -0.43 eV, is compared with the experimental spectrum by Bazs\'o et al.~\cite{bazso_tautomers_2011}. The grey region is the energy band from where initial conditions were chosen. The FWHM employed to convolute the spectra is 0.15 eV.}
  \label{fig:spectra}
\end{figure}

In Table~\ref{tab:exc}, spin-free excitation energies and oscillator strengths of both tautomers calculated at the SA10S+10T-CASSCF(12,9)/6-31G* level of theory are given. The excitation energies obtained are slightly higher than those reported experimentally and than those calculated at a more correlated level of theory. In any case, the state ordering of the lowest singlet states agrees with that predicted both by DFT/MRCI\cite{tomic_quantum_2005} as well as by MRCI.\cite{barbatti_photodynamical_2011}
The brightest state in both tautomers is the $S_1$, corresponding to a $\pi\pi^*$ excitation. The remaining calculated singlet states are dark at the Franck-Condon geometry and correlate with $n\pi^*$ transitions. Note that at the Franck-Condon point the $n_O$ and $n_N$ orbitals are mixed and thus we denote the transition simply as $n\pi^*$. At other geometries (as discussed below) these orbitals do not mix anymore and the particular $n$ orbital ($n_O$ or $n_N$) will be specified.
Since the $S_4$ in the \emph{keto} and the $S_3$ in the \emph{enol} form are well separated from the lower states, singlet states $S_{\geq4}$ and $S_{\geq3}$, respectively, were not considered for the dynamics simulations.
For completeness, the lowest triplet states at the equilibrium geometry are also reported in Table~\ref{tab:exc}. In both tautomers, the two lowest triplet states, $T_1$ and $T_2$, correspond to $\pi\pi^*$ while the $n\pi^*$ is the $T_3$ state. Again based on the energetic separation, states above $T_3$ in the \emph{keto} and $T_4$ in the \emph{enol} tautomer were not included in the dynamics simulations.

The SA10S-CASSCF(12,9)/6-31G* spectra based on 2000 (1000) geometries of the \emph{keto} (\emph{enol}) tautomer are displayed in Fig.~\ref{fig:spectra}a (b). The first band of the spectrum of the \emph{keto} tautomer (Fig.~\ref{fig:spectra}a) results mainly from four singlet excited states while three are the most important in the first band of the \emph{enol} spectrum (Fig.~\ref{fig:spectra}b). The \emph{keto} spectrum shows a much broader absorption band than the \emph{enol} one. This might be a hint at the larger excited-state gradients in the Franck-Condon region of \emph{keto}-cytosine.
Noteworthy is that the contributions from both $S_1$ and $S_2$ to the \emph{keto} spectrum are comparably large, indicating that the $\pi\pi^*$ and $n\pi^*$ states are close in the Franck-Condon region and the state ordering may be easily inverted for different geometries within this region.
% perhaps that the structure of the keto spectrum is artificial
In the \emph{enol} spectrum, the $S_1$ contributes strongest and it can be inferred that this lowest excited state corresponds to the bright $\pi\pi^*$ state at most of the geometries.

Figure~\ref{fig:spectra}c shows the experimental spectrum along with a linear combination of the simulated spectra, where the ratio for the contribution of the respective tautomer is 0.24:0.76 (\emph{keto}:\emph{enol}). This ratio corresponds to the one in Ref.~\onlinecite{bazso_tautomers_2011} when ignoring the \emph{imino} tautomer and assuming that both \emph{enol} rotamers yield the same spectrum. Since the CASSCF excitation energies are overestimated, the simulated spectrum was shifted by 0.43~eV to obtain the maximum overlap with the experiment.
%Because the \emph{enol} tautomer shows both a higher abundance and a higher absorption in the considered energy range (compare Figs.~\ref{fig:spectra}a and b), it strongly dominates the first band of the total absorption spectrum. Moreover, since the oscillator strength of the $\pi\pi^*$ state of the \emph{enol} tautomer (see Table~\ref{tab:exc}) is too small at the CASSCF level of theory (compare to e.g. 0.1389 from DFT/MRCI\cite{tomic_quantum_2005}), the total spectrum should be even more dominated by the \emph{enol} form. (In passing, we note that too small oscillator strengths have no impact on the dynamics simulations.)
The \textit{keto} tautomer shows a stronger absorption in the considered energy range (compare Figs.~\ref{fig:spectra}a and b), as a direct consequence that the $\pi\pi^*$ state of the keto is brighter than the one of the enol (see Table~\ref{tab:exc}). Still, the contributions of the \textit{keto} and \textit{enol} forms to the total spectrum are comparable due to the higher abundance of the \textit{enol} form. Moreover, since the oscillator strength of the $\pi\pi^*$ state of the \emph{enol} tautomer (see Table~\ref{tab:exc}) is too small at the CASSCF level of theory (compare to e.g. 0.1389 from DFT/MRCI\cite{tomic_quantum_2005}), the total spectrum should be more dominated by the \emph{enol} form. (In passing, we note that too small oscillator strengths have no impact on the dynamics simulations.)
The composite spectrum (Figure~\ref{fig:spectra}c) was generated mainly to assess the energy shift between CASSCF and experiment so that a proper excitation range for the initial condition generation could be chosen. The agreement between the experimental spectrum and the shifted calculated one is otherwise acceptable, considering the level of theory employed.

\subsection{Ionization processes}\label{ssec:ionization}

\begin{figure}
  \centering
  \includegraphics[width=246pt]{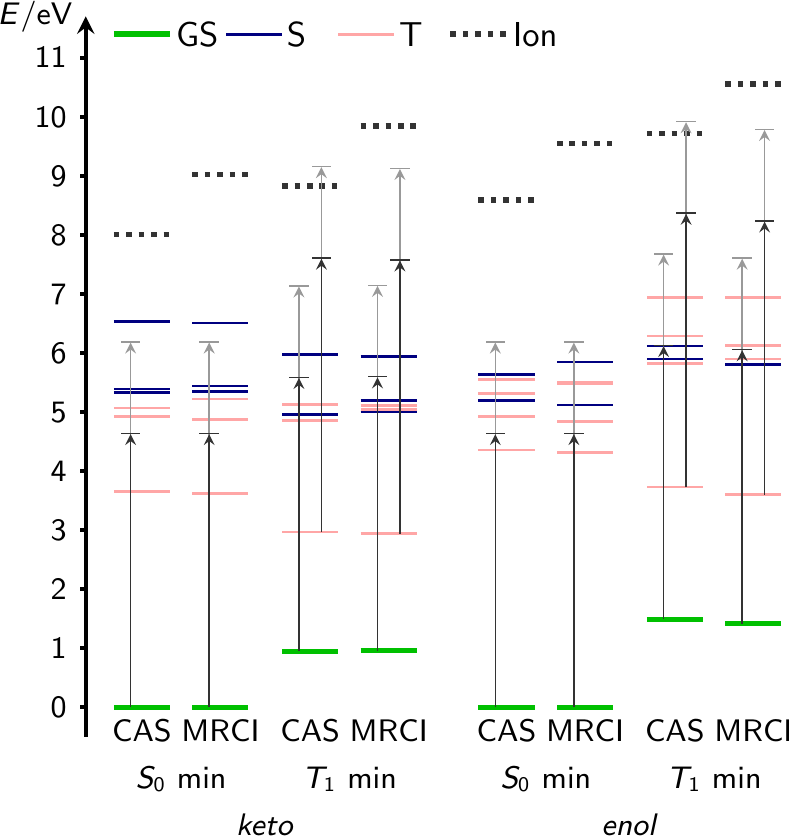}
  \caption{Neutral and ionic CASSCF(12,9)/6-31G* (\emph{keto}: SA4S+1D+3T, \emph{enol}: SA3S+1D+4T) and MRCI+Q energies at the $S_0$ and $T_1$ minima (optimized at the same CASSCF level) for the \emph{keto} form (left) and the \emph{enol} form (right). Colors denote ground state (GS), excited singlets (S), triplets (T) and the lowest ionic state (Ion). The vertical arrows indicate the excitation energy of a 3$\times$800 nm (4.6 eV; black) and 2$\times$400 nm (6.2 eV; grey) probe laser.}
  \label{fig:ionization}
\end{figure}

In the following, we discuss how excited-state relaxation can be detected experimentally and how experimental results can be related to our simulations.

% keto: dark triplet
In gas-phase ultrafast time-resolved experiments, the excited-state populations are usually detected by means of photoionization. Thus, in order to detect a signal, the energy difference between the ionic state and the populated neutral state (the ionization potential) needs to be smaller than the energy provided by the probe laser.
Most of the time-resolved studies~\cite{kang_intrinsic_2002,kosma_excited-state_2009,ho_disentangling_2011} on cytosine use a three-photon 800 nm probe, which corresponds to ca. 4.6 eV. Some experiments use two-photon 400 nm\cite{canuel_excited_2005} or 200 nm\cite{ullrich_electronic_2004} probe pulses instead, which is equivalent to 6.2 eV. In all cases, at the Franck-Condon region all but the ground state is supposed to be detected, so that it is assumed that the time constants measured experimentally correspond exclusively to the relaxation of the excited population to the ground state. The latter assumption includes that all triplet states are also ionized by the probe pulses or that triplet state population is negligible. As we show below, this assumption might not necessarily always be true.

% figure: ionization
Figure~\ref{fig:ionization} shows the energies of the singlet and triplet states considered in the dynamics as well as the lowest ($N$-1)-electron (ionic) state at the $S_0$ and $T_1$ minima for the \emph{keto} and \emph{enol} tautomers, calculated at the CASSCF(12,9)/6-31G* (\emph{keto}: state-averaging over 4 singlets, 1 doublet and 3 triplets, denoted as SA4S+1D+3T; \emph{enol}: SA3S+1D+4T)
%state-averaging as in the dynamics, see section~\ref{ssec:abinitio})
and the internally-contracted MRCI+Q\cite{KW88,WK88,KW92} level of theory (Q indicating Davidson correction). In the MRCI, all inner shells were kept frozen (8 orbitals) and only doubly external excitation were considered in order to keep the calculations at a reasonable computational cost.
Arrows in black and grey indicate probe excitation energies of 4.6 eV (3$\times$800 nm) and 6.2 eV (2$\times$400 nm), respectively.
As it can be seen, the CASSCF and MRCI+Q energies for the neutral excited states are very similar, which justifies performing CASSCF dynamics. On the contrary, the energy of the corresponding ionic state at the MRCI+Q level of theory is strongly destabilized compared with the CASSCF energy. The MRCI+Q values are in good agreement with photoelectron measurements\cite{Dougherty1977JCP,Yu1978JACS} and previous calculations.\cite{Dolgounitcheva2003JPCA} According to the more reliable MRCI+Q energies, at the $S_0$ minima of both tautomers, the ground state indeed cannot be ionized by any of the mentioned probe pulses. These geometries correspond to the starting point of the relaxation dynamics and it is thus unimportant that the energy of the probe laser suffices to ionize the $T_1$ since it is not (yet) populated.
%the $T_1$ state can be ionized in some experimental setups since it will not (yet) be populated.
However, all initially populated excited singlet states can be ionized at the $S_0$ minimum geometry.

As shown by Richter \textit{et al.}\cite{richter_femtosecond_2012} and in the present study (see below), the triplet states play a significant role in the relaxation dynamics of cytosine. It is therefore justified to look at the ionization potential at the endpoint of a triplet relaxation pathway: the $T_1$ minimum. At this geometry, the energy of the $T_1$ state is lowered while the ionic state is considerably destabilized. Accordingly, ionization from the $T_1$ minimum should be negligible with the energies of the mentioned experimental setups (see MRCI+Q values) and we expect the population that has flown from the excited singlets to the $T_1$ to be experimentally hardly distinguishable from the population having relaxed to the ground state.
Thus, we propose that the transients observed experimentally arise from both the relaxation to the ground state \emph{and} to the $T_1$.
%Note however, that using a 193 nm probe laser Nir et al.\cite{nir_rempi_2002} assigned ns timescale to long-lived triplet states.

%However, with multiphoton setups, ionization is the less probable the higher the ionization potential (because more photons are needed).

% =========================================================================================================================================== %

\subsection{Excited state lifetimes}\label{ssec:lifetimes}

\begin{figure}
  \centering
  \includegraphics[width=246pt]{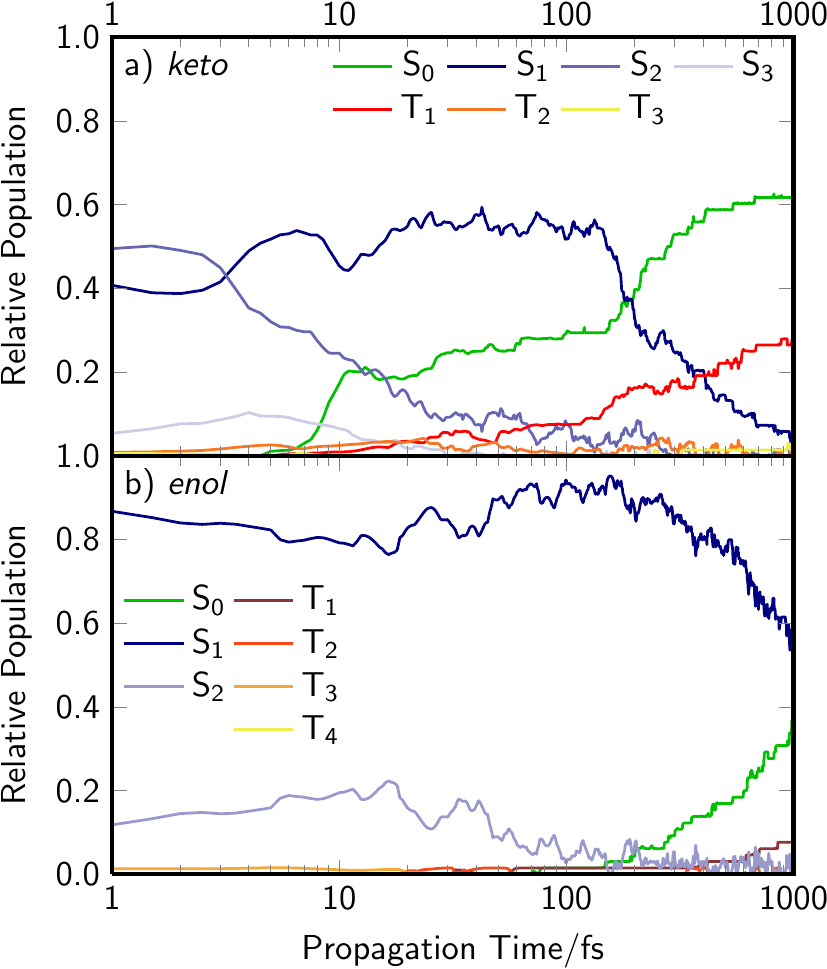}
  \caption{Average quantum amplitudes of the considered states in the dynamics of the \emph{keto} (a) and the \emph{enol} (b) forms of cytosine. }
  \label{fig:population}
\end{figure}

% general
The time-dependent populations of the electronic states are given by the squares of the coefficients $|c_\alpha^{\text{MCH}}|^2$, where
\begin{equation}
  \mathbf{c}^{\text{MCH}}(t)=\mathbf{U}(t)\mathbf{c}^{\text{diag}}(t).
\end{equation}
These values are presented in Figure~\ref{fig:population}a and \ref{fig:population}b for the \emph{keto} and the \emph{enol} forms, respectively. As can be clearly seen, the excited-state dynamics of the two tautomers differ dramatically.

% keto, main
In the \emph{keto} tautomer, we observe an ultrafast internal conversion (IC) from the $S_2$ to the $S_1$ and also to $S_0$, in agreement with previous studies.~\cite{gonzalez-vazquez_time-dependent_2010,barbatti_photodynamical_2011,richter_femtosecond_2012} About 20\% of all trajectories return to the ground state in about 10~fs. Within 500~fs, more than 60\% return to the ground state. On the same timescale, the $T_1$ is populated as a result of the deactivation cascade $S_1\rightarrow T_2\rightarrow T_1$ by about 25\% of all trajectories (see also subsection~\ref{ssec:keto} below). A small fraction of the population remains in the excited singlet states for more than 1 ps. In comparison to the other states, the $S_3$ and $T_3$ play a minor role in the deactivation mechanism.

% enol, main
The dynamics of \emph{enol} cytosine is completely different (see Fig.~\ref{fig:population}b). In this case, the $S_1$ is populated by more than 80\% in the beginning and by more than 90\% after 100~fs. Compared to the \emph{keto} form, relaxation to the ground state is much slower (the first trajectory relaxes after 150 fs and after 1 ps only 30\% of all trajectories are relaxed). Also differently from the \emph{keto} form, ISC is considerably less important (ca. 5\% in the \emph{enol} versus ca. 25\% in the \emph{keto} form).

% keto: fit
Based on the discussion of subsection~\ref{ssec:ionization}, we now present fits of the population decay to the experimentally unobservable states: the $S_0$ and the $T_1$. Based on the $S_0$ population of Fig.~\ref{fig:population}a, we anticipate a biexponential decay to the two mentioned states in the \emph{keto} tautomer:
\begin{equation}
  f(t)=c\left(1-\gamma_1\cdot\text{e}^{-t/\tau_1}-(1-\gamma_2)\cdot\text{e}^{-t/\tau_2}\right),\label{eq:bi}
\end{equation}
while the \emph{enol} tautomer can be treated monoexponentially:
\begin{equation}
  f(t)=1-\text{e}^{-t/\tau_3}.\label{eq:mono}
\end{equation}

The time constants and fitting parameters according to equations \eqref{eq:bi} and \eqref{eq:mono} are collected in Table \ref{tab:fit}. Since after 1 ps in the \emph{keto} tautomer a small fraction of the trajectories is still in the $S_1$, we additionally performed a triexponential fit. In the latter, $\tau_1$ and $\tau_2$ remain almost unchanged with respect to the biexponential fit, while the third time constant ($\tau_3$, not to be confused with $\tau_3$ in Eq.~\eqref{eq:mono}) is quite large and also has a large uncertainty (2200$\pm$1700 fs). For completeness, also time constants for the $S_0$ and $T_1$ are given separately. For the $T_1$ in the \emph{enol} tautomer, the simulation time of 1 ps was not sufficient to extract any meaningful time constant.

\begin{table}
  \centering
  \caption{Time constants and fitted parameters.}
  \begin{tabular}{lcccccc}
    \hline
    State       &$c$    &$\tau_1$ (fs)  &$\gamma_1$     &$\tau_2$ (fs)  &$\gamma_2$     &$\tau_3$ (fs)\\
    \hline
    \multicolumn{7}{c}{--- \emph{keto} ---}\\
    $S_0$       &0.658  &6              &0.196          &230            &--             &--\\
    $T_1$       &0.322  &4              &0.110          &350            &--             &--\\
    $S_0+T_1$   &0.952  &7              &0.161          &270            &--             &--\\
    $S_0+T_1$   &\emph{1.000}$^a$  &7       &0.152          &260            &0.774          &2200\\
    \hline
    \multicolumn{7}{c}{--- \emph{enol} ---}\\
    $S_0$       &--  &--      &--             &--             &--             &2400\\
    $S_0+T_1$   &--  &--      &--             &--             &--             &1900\\
    \hline
  \end{tabular}
  \label{tab:fit}
\\
$^a$Value of $c$ fixed to 1.0 in the fit.
\end{table}

% both: table
As discussed above, we believe that most experimental time constants should relate to a combined $S_0$+$T_1$ relaxation pathway and therefore focus on our $S_0$+$T_1$ fits henceforth. Table~\ref{tab:comp} summarizes these lifetimes and also contains time constants reported in the literature, both experimentally and theoretically. As it can be seen, the time constants of the present work are in very good agreement with the experimental results, especially given the broad range of the latter. According to our simulations, we assign the faster lifetimes $\tau_1$ and $\tau_2$ to the \emph{keto} tautomer and the slower $\tau_3$ to the \emph{enol} form. Such a distinction between the tautomers is difficult in the experiments. Due to the employed pulse durations, none of the experimental studies is able to accurately resolve the shortest time constant $\tau_1$, which we determine as 7 fs. However, values of $<$100 fs\cite{kosma_excited-state_2009} or 50 fs\cite{ullrich_electronic_2004,kotur_following_2012} are given in the literature. The second calculated time constant $\tau_2$ (270 fs)
%matches the experimental values ranging from 160 fs\cite{canuel_excited_2005} to 820 fs.~\cite{ullrich_electronic_2004}
falls within the range of experimental values from 160 fs\cite{canuel_excited_2005} to 820 fs.~\cite{ullrich_electronic_2004}
Similar time constants also have been reported in previous dynamics simulation studies.\cite{lan_photoinduced_2009,hudock_excited-state_2008,barbatti_photodynamical_2011} The observed mechanistic details differ considerably in these studies since the employed level of theory for the on-the-fly calculations is different (vide infra). Also the third time constant $\tau_3$ from the enol at 1.9 ps agrees with the experimental findings. Kotur et al.\cite{kotur_distinguishing_2011,kotur_following_2012,kotur_phd_2012} attributed a comparable time constant (2.3 ps) to the relaxation of the \emph{keto} tautomer. Similar to what was done in Ref.~\onlinecite{barbatti_photodynamical_2011}, we can also obtain a ps lifetime in the keto form (2200~fs) so that relaxation of this tautomer on this timescale cannot be completely ruled out. However, we believe that the experimentally observed timescale of a few ps should be attributed primarily to the \emph{enol} tautomer, based on its higher relative abundance. This assignment is in line with ultrafast experiments conducted with lower pump energies,\cite{kosma_excited-state_2009,ho_disentangling_2011} in which the \emph{enol} tautomer is not excited and consequently the slow (ps) component vanishes.

\begin{table}
  \centering
  \caption{Time constants obtained in this work compared with values from the literature. According to the results presented in Fig. \ref{fig:ionization}, we assume relaxation to both $S_0$ and $T_1$ to be captured in the experimental time constants.}
  \begin{tabular}{llccc}
    \hline
    Study            &Setup                                     &$\tau_1$ (fs)           &$\tau_2$ (fs)          &$\tau_3$ (fs)\\
    \hline
    \multicolumn{5}{c}{--- \emph{This work} ---}\\
    \emph{keto}             &$S_0$+$T_1$                        &7                       &270                    &(2200)$^a$     \\
    \emph{enol}             &$S_0$+$T_1$                        &--                      &--                     &1900      \\
    \multicolumn{5}{c}{--- \emph{Experimental} ---}\\
    Ref. \onlinecite{kang_intrinsic_2002}                       &$n\times$800 nm &--     &--                    &3200     \\
    Ref. \onlinecite{canuel_excited_2005}                       &2$\times$400 nm &--     &160                   &1860     \\
    Ref. \onlinecite{ullrich_electronic_2004}                   &200 nm          &$<$50  &820                   &3200     \\
    Ref. \onlinecite{kosma_excited-state_2009}                  &3$\times$800 nm &$<$100 &210                   &2200  \\
    Ref. \onlinecite{ho_disentangling_2011}                     &3$\times$800 nm &--     &500                   &4500  \\
    Ref. \onlinecite{kotur_distinguishing_2011},\onlinecite{kotur_following_2012}       &$n\times$780 nm &50     &240           &2360  \\
    \multicolumn{5}{c}{--- \emph{Theoretical} ---}\\
    Ref. \onlinecite{lan_photoinduced_2009}                     &SH: OM2/MRCI    &40     &370                   &--\\
    Ref. \onlinecite{hudock_excited-state_2008}                 &AIMS: CAS(2,2)  &$<$20  &$\sim$800             &--\\
    Ref. \onlinecite{barbatti_photodynamical_2011}              &SH: CAS(14,10)  &9      &527                   &3080\\
    \hline
  \end{tabular}
    \label{tab:comp}
\\
$^a$The value in parenthesis correspond to the triexponential fit reported in Table II.
\end{table}

\begin{figure}
 \centering
 \includegraphics[width=246pt]{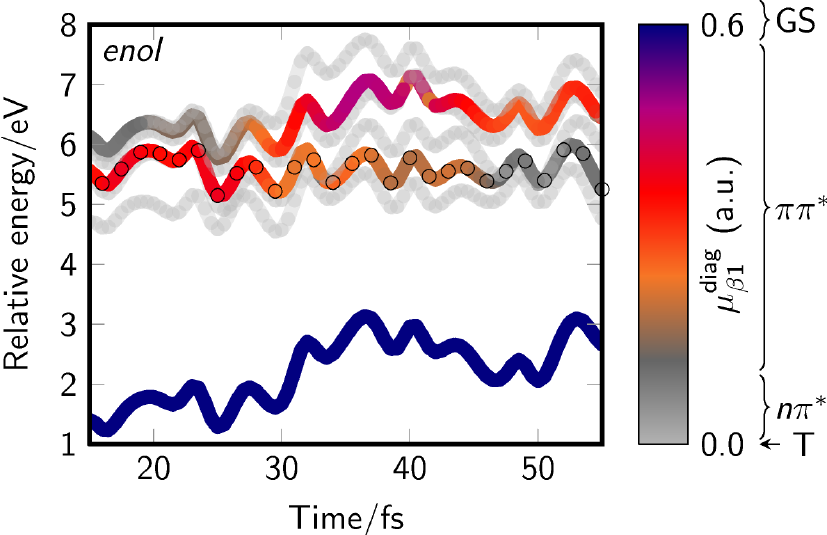}
 \caption{Part of an exemplary trajectory of \emph{enol} cytosine moving on the $S_1$ surface (circles). Overlapped on the potential energy of the different states is the transition dipole moment $\mu_{\beta0}^{\text{diag}}$  given by the color coding of the curves. Very small transition dipole moments (light grey) correspond to triplet states. Values of $\mu_{\beta0}^{\text{diag}}$ above 0.6 a.u. indicate the ground state (blue). The excited singlet states given by the intermediate colors, where the $n\pi^*$ state is identified by values below 0.1 a.u. (dark grey) and the $\pi\pi^*$ state by values between 0.1 and 0.6 a.u. (shades of red).}
 \label{fig:trajectory}
\end{figure}

\begin{table}
  \centering
  \caption{Total transition dipole moment $|\mu_{\beta0}^{\text{diag}}|^2$ (a.u.) used to classify the classically occupied state.}
  \begin{tabular}{lcccc}
    \hline
    Tautomer            &Ground state   &$\pi\pi^*$     &$n\pi^*$       &Triplet\\
    \hline
    \emph{keto}       &$>1.3$   &$>0.10$   &$>10^{-6}$ &$>0.0$\\
    \emph{enol}       &$>0.6$   &$>0.13$  &$>10^{-6}$ &$>0.0$\\
    \hline
  \end{tabular}
  \label{tab:class}
\end{table}

% model: mixture, ratios, classification, fits

In order to better explain the spectroscopic observations, which do not only depend on the populations of the electronic states but also on transition dipole moments, an analysis of the involved excited states has been performed.
%Besides the simple fit procedure, we also performed a more complex analysis in order to better explain the spectroscopic observations, which do not only depend on the populations of the electronic states but also on transition dipole moments.
Spectroscopic results are often discussed in terms of diabatic states, where the wavefunction character and thus properties like the transition dipole moments change as little as possible. However, our trajectories  are calculated in the basis of the eigenfunctions of a Hamiltonian operator and these eigenfunctions change the wavefunction character. Since the diabatic (``spectroscopic'') state populations cannot be obtained from our simulations, we used the transition dipole moments to calculate approximate spectroscopic populations as explained below.

For all trajectories, the occupied state was classified as ground state (GS), $\pi\pi^*$, $n\pi^*$ or triplet (T) on the basis of the transition dipole moment $\mu_{\beta0}^{\text{diag}}$ between the state where the trajectory is moving and the state lowest in energy (transformed into the diagonal basis, see section \ref{sec:methodology}).
Table \ref{tab:class} gives the magnitudes of the transition dipole moments used for this classification. Note that these values are based on a qualitative examination of the trajectories and thus the assignment cannot be considered strict. Figure~\ref{fig:trajectory} shows  a snapshot of the potential energies vs. time of one exemplary trajectory of the \emph{enol} tautomer, where the trajectory starts on the lowest excited singlet surface. The curves are colored according to the magnitude of the transition dipole moment $\mu_{\beta0}^{\text{diag}}$;
note that $\mu_{00}^{\text{diag}}$ corresponds to the permanent dipole moment of the electronic ground state.
As it can be seen, the magnitude of $\mu_{\beta0}^{\text{diag}}$ is a convenient indicator of the wavefunction character. Triplet states show virtually no transition dipole moments and are given in light grey. We choose comparably small values of $\mu_{\beta0}^{\text{diag}}$ (see Tab. \ref{tab:class}) to be attributed to dark $n\pi^*$ states (dark grey), while the bright $\pi\pi^*$ state is identified by comparably large values (given as shades of red). With this classification it can be seen how this particular trajectory is initially in a state with $\pi\pi^*$ character and after 50 fs has continuously evolved to a dark $n\pi^*$ configuration, while one of the upper states becomes the bright $\pi\pi^*$ state.

\begin{figure}
  \centering
  \includegraphics[width=246pt]{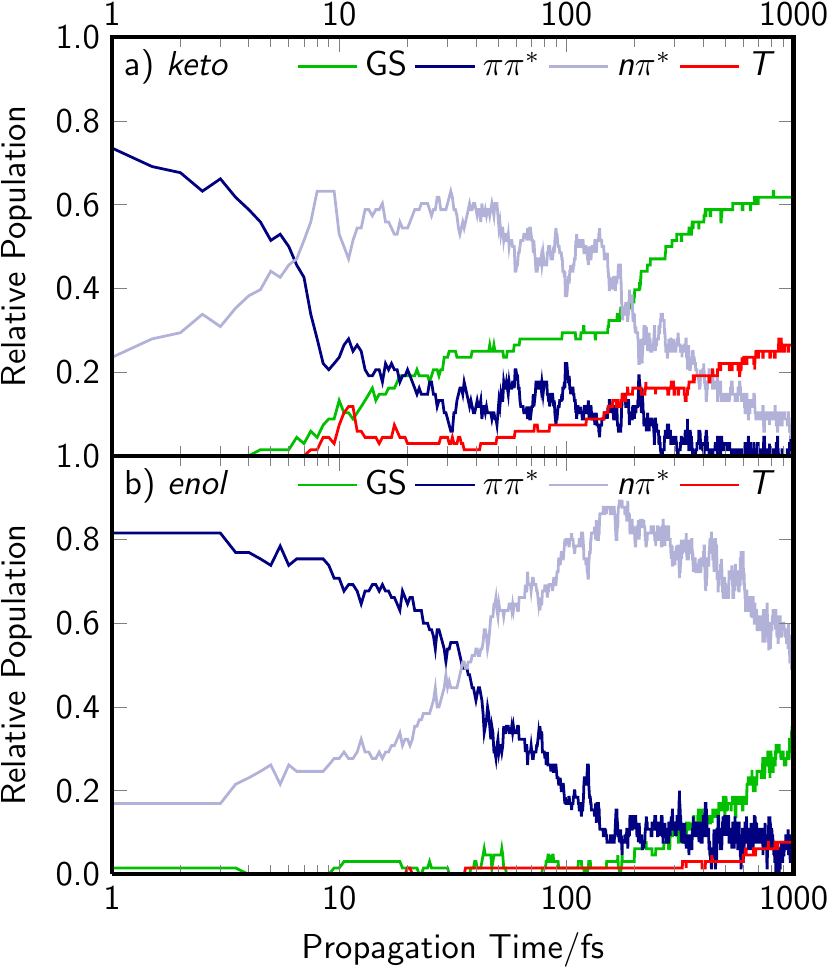}
  \caption{Approximate spectroscopic state populations of the \emph{keto} (a) and \emph{enol} (b) forms of cytosine. }
  \label{fig:diab}
\end{figure}

% both
Using the information obtained from the transition dipole moments, the evolution of the number of trajectories in each of the defined ``spectroscopic'' classes (GS, $n\pi^*$, $\pi\pi^*$ and T), which we will refer to as populations of the spectroscopic states, are depicted in Fig.~\ref{fig:diab}. Comparing with the populations in Fig.~\ref{fig:population}, the ground state and triplet states are well identified, since the spectroscopic and the actual populations of these states are roughly the same.

% keto
For the \emph{keto} tautomer, one should be careful not to make a correspondence between the states $S_1$ and $S_2$ in the MCH basis (recall Fig.~\ref{fig:spectra}a) and the spectroscopic states $\pi\pi^*$  and $n\pi^*$. As already discussed, $S_1$ and $S_2$ are close in energy at the Franck-Condon region and since this region comprises a range of geometries, $S_1$ and $S_2$ may strongly mix and thus both exhibit $n\pi^*+\pi\pi^*$ character. This mixing together with the uncertainty of the classification scheme explains the initial population of the $n\pi^*$ state given in Fig.~\ref{fig:diab}.

% enol
For the \emph{enol} tautomer, the assignment of the states $S_1$ and $S_2$ clearly changes over time. While in the beginning $S_1$ mostly corresponds to $\pi\pi^*$ and $S_2$ to $n\pi^*$, this is quickly reversed with a time constant of approximately 40~fs, suggesting an adiabatic change of the wavefunction character. Since it has been shown before\cite{spanner_dyson_2012,kotur_neutral-ionic_2012} that the excited-state wavefunction character may strongly influence ionization yields, the $\pi\pi^*\rightarrow n\pi^*$ conversion is a possible explanation for the shortest time constant observed in the various experiments.\cite{kosma_excited-state_2009,ho_disentangling_2011, kotur_following_2012}
%the 40 fs time constant of the $\pi\pi^*\rightarrow n\pi^*$ conversion is possibly observed in the experiments.\cite{kosma_excited-state_2009,ho_disentangling_2011, kotur_following_2012} }

% =========================================================================================================================================== %

\subsection{Relaxation mechanism of keto cytosine}\label{ssec:keto}

\begin{figure}
  \centering
  \includegraphics[scale=1]{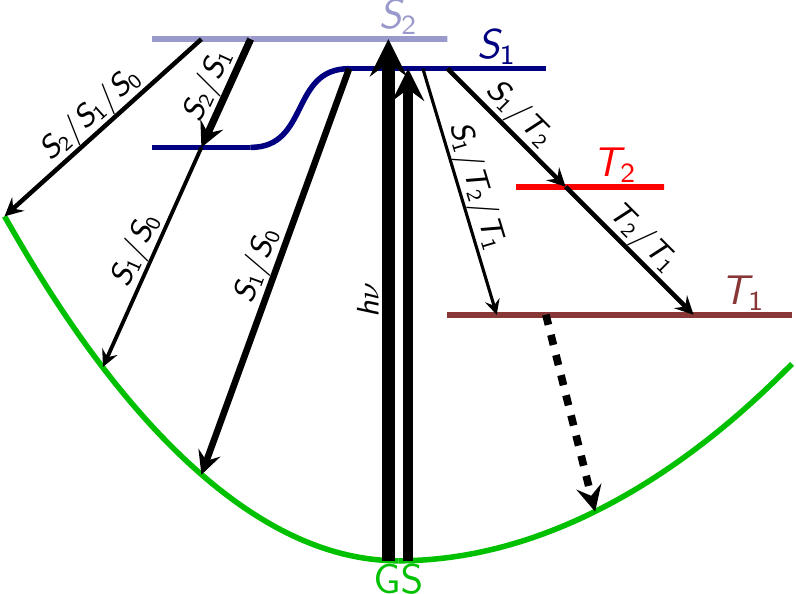}
  \caption{Schematic representation of the relaxation pathways observed in \emph{keto} cytosine after excitation (indicated by $h\nu$). Arrow thickness depicts the fraction of trajectories taking a particular pathway. ISC from $T_1$ to $S_0$ is not observed in the present work, only assumed; this is indicated by a dashed line.}
  \label{fig:overview_keto}
\end{figure}

The dynamics simulations of the \emph{keto} tautomer show ultrafast IC processes, as found in previous studies.\cite{lan_photoinduced_2009,hudock_excited-state_2008, gonzalez-vazquez_time-dependent_2010,barbatti_photodynamical_2011, richter_femtosecond_2012} Moreover, we observe ISC processes, which will be discussed below. The different relaxation pathways are summarized in Figure~\ref{fig:overview_keto}.

\begin{figure}
  \centering
  \includegraphics[scale=1]{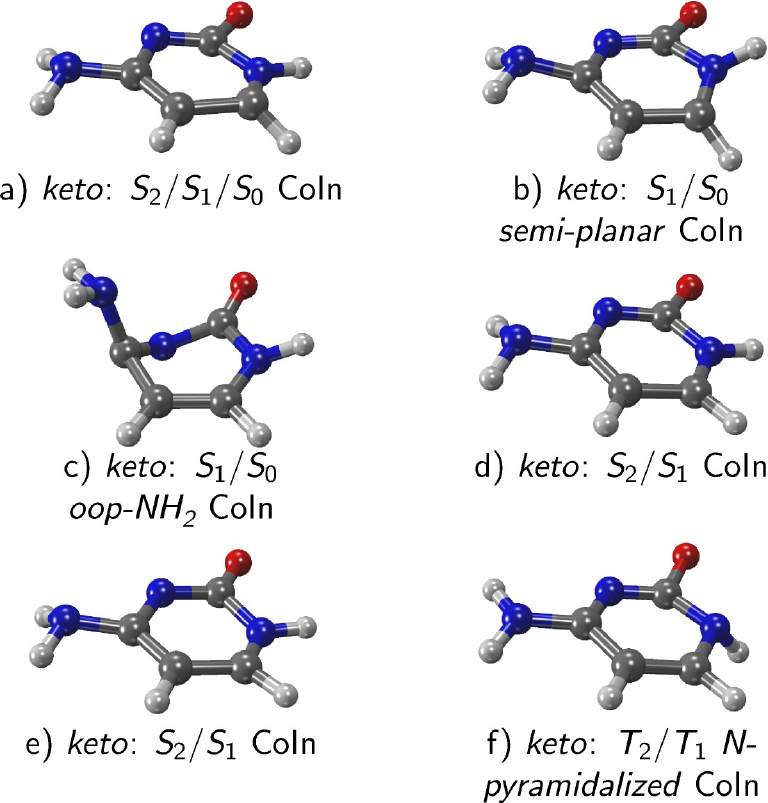}
  \caption{CoIns of the \emph{keto} tautomer optimized at the SA4S+3T-CASSCF(12,9)/6-31G* level of theory. The three-state $S_2/S_1/S_0$ structure is optimized as a two-state CoIn between $S_2$ and $S_0$. The coordinates of all geometries are given in the Supporting Information.}
  \label{fig:geom_keto}
\end{figure}

We observed three different relaxation cascades from the $S_1$ and $S_2$ excited states to the $S_0$. The population in $S_1$ is deactivated directly ($S_1\rightarrow S_0$), while that in $S_2$ either relaxes by first decaying to the $S_1$ and later to the $S_0$ ($S_2\rightarrow S_1\rightarrow S_0$) or directly via a three-state CoIn ($S_2\rightarrow S_0$).~\cite{gonzalez-vazquez_time-dependent_2010}
These relaxation cascades employ a number of CoIns, which are depicted in Figure~\ref{fig:geom_keto}a-e. Out of all trajectories decaying to the ground state during the course of the \emph{keto} simulation, the ones giving rise to the time constant $\tau_1$ (25\% of all trajectories) relax exclusively via two CoIns: the three-state $S_2/S_1/S_0$ CoIn (Fig.~\ref{fig:geom_keto}a) and the $S_1/S_0$ \emph{semi-planar} CoIn (Fig.~\ref{fig:geom_keto}b). The labelling of the two-state CoIns follows the nomenclature of Ref. \onlinecite{barbatti_photodynamical_2011}. Both the three-state $S_2/S_1/S_0$ CoIn~\cite{blancafort_key_2004,kistler_three-state_2008,gonzalez-vazquez_time-dependent_2010}  and the $S_1/S_0$ \emph{semi-planar} CoIn\cite{ismail_ultrafast_2002, hudock_excited-state_2008, merchan_ultrafast_2003, blancafort_singlet_2005, blancafort_energetics_2007, kistler_radiationless_2007, barbatti_photodynamical_2011}  have been reported by several groups previously.
Compared to the ground state geometry, both CoIns are characterized primarily by an elongated \ce{C2=O} and \ce{C5-C6} bond and a compressed \ce{C2-N3} bond while mostly retaining ring planarity (see Fig.~\ref{fig:tautomers}a for atom labeling).
% Since they also exhibit basically the same wavefunction character of the crossing states, they can be thought of as lying on the same seam.

The $S_1/S_0$ \emph{semi-planar} CoIn is also employed by \emph{keto} cytosine to relax to the $S_0$ at later times, contributing to $\tau_2$ (17\% of all trajectories).
The remaining trajectories (20\%) returning to the $S_0$ do so via another $S_1/S_0$ CoIn, the \emph{oop-\ce{NH2}} CoIn (Fig.~\ref{fig:geom_keto}c), which shows a semi-twisted ring structure with puckering at \ce{N3} and \ce{C4} and a strong out-of-plane distortion of the amino-group.
Only one trajectory relaxed through the so-called \emph{\ce{C6}-puckered} $S_1/S_0$ CoIn,\cite{barbatti_photodynamical_2011} which was therefore not optimized. %and thus it was not optimized.
% \textbf{The branching ratios are best comparable to the values of Barbatti.\cite{barbatti_photodynamical_2011}\tdo{???}}
Rapid interconversion between $S_2$ and $S_1$ (usually in less than 30 fs) is facilitated by two more CoIns, given in Fig.~\ref{fig:geom_keto}d and~\ref{fig:geom_keto}e. Both CoIns were reported by Kistler et al.,\cite{kistler_radiationless_2007} with the names  $R_x(ci12)$  (Fig.~\ref{fig:geom_keto}d) and
$R_x(ci12)^\prime$ (Fig.~\ref{fig:geom_keto}e). Based on our dynamics simulations, $R_x(ci12)$ (Fig.~\ref{fig:geom_keto}d) seems to be the major funnel for $S_2\rightarrow S_1$ interconversion.

The deactivation mechanism described above is similar to the one obtained by Barbatti et al.~\cite{barbatti_photodynamical_2011} using CASSCF(14,10) in the singlet manifold only. Their simulations also find the $S_1/S_0$ \emph{semi-planar} CoIn to be important in the early (16\%) and late (52\%) deactivation from the $S_1$.  An explanation why in our simulations this CoIn is less accessed at a later time (only 17\%) is that part of the population is transferred to the triplet states. Additionally, Barbatti et al.~\cite{barbatti_photodynamical_2011} observe the \emph{oop-\ce{NH2}} (7\%) and the \emph{\ce{C6}-puckered} CoIn (8\%) $S_1/S_0$ CoIns, while we only find the former, probably because of the level of theory employed. The multiple spawning dynamics of 
Hudock and Mart\'{\i}nez\cite{hudock_excited-state_2008} based on CASSCF(2,2), in contrast, finds the \emph{oop-\ce{NH2}} CoIn to be the most important deactivation channel (65\%), while the semiempirical surface-hopping dynamics simulations at the OM2 level of theory of Lan and coworkers~\cite{lan_photoinduced_2009} only see deactivation to the $S_0$ state via the \emph{\ce{C6}-puckered} CoIn. Fair to say, however, is that all the electronic structure calculations in the dynamical simulations up to date do not include dynamical correlation (e.g. at CASPT2 level of theory or MRCI), which could in principle change the shape of the PESs. That said, the CASPT2 calculations of Blancafort~\cite{blancafort_energetics_2007} indicate that the \emph{\ce{C6}-puckered} CoIn is more accessible than what is expected at CASSCF level of theory, see also Ref.~\cite{merchan_unified_2006} Similarly, the structure optimized for the $S_1/S_0$ \emph{semi-planar} CoIn is higher in energy when dynamical correlation is included\cite{blancafort_energetics_2007,kistler_three-state_2008,blancafort_key_2004} which implies that the time scales obtained for the $\tau_1$ in the present work and the other dynamical studies~\cite{hudock_excited-state_2008,lan_photoinduced_2009,barbatti_photodynamical_2011} could be underestimated.

\begin{figure}
  \centering
  \includegraphics[scale=1]{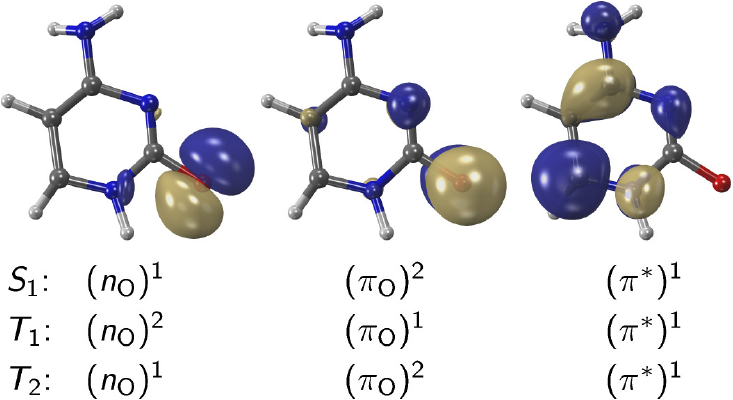}
  \caption{Orbitals at the ISC geometries and predominant configurations of $S_1$, $T_1$ and $T_2$ at these geometries.}
  \label{fig:orb_keto}
\end{figure}

One important finding in our simulations is that ISC in \emph{keto} cytosine is ultrafast and competes with IC, in agreement with the previous simulations of Richter et al.\cite{richter_femtosecond_2012} However, while in the latter study most ISC hops were observed already after 10 fs, here ISC takes place on a timescale of few hundreds of fs. This slower ISC is because propagating in the MCH basis (see Section IIA) makes the algorithm much more stable in the case of weakly coupled surfaces, as it is the case for small spin-orbit couplings (SOCs).  The here predicted ISC time scale is slower than the $S_2\rightarrow S_1$ conversion and therefore ISC only occurs from the lowest singlet surface.
Despite the minor differences with respect to the findings of Richter et al.,\cite{richter_femtosecond_2012}, this work confirms that the main ISC path is $S_1\rightarrow T_2\rightarrow T_1$ and that the involved triplet IC ($T_2\rightarrow T_1$) is extremely fast.

As depicted in Fig.~\ref{fig:overview_keto}, a second ISC path is the direct $S_1\rightarrow T_1$ transition. However, as both processes (direct $S_1\rightarrow T_1$ and indirect $S_1\rightarrow T_2\rightarrow T_1$) employ the same $S_1/T_2/T_1$ near-degeneracy, the two pathways are mechanistically very similar.
Interestingly, efficient ISC proceeds via a three-state near-degeneracy of the $S_1$, $T_2$ and $T_1$ states and not at singlet-triplet crossings involving only two states.
%at $S_1/T_1$ crossing points.
$S_1$ and $T_2$ have mainly $n_O\pi^*$ character (see Fig.~\ref{fig:orb_keto}) at the points where both states cross.
%The need for a three-state near-degeneracy for efficient ISC stems from the fact that $S_1$ and $T_2$ are mainly of $n_O\pi^*$ character (see Fig.~\ref{fig:orb_keto}) at the points of the crossing of both states.
In accordance with the El-Sayed rule,\cite{elsayed_spinorbit_1963} SOCs between the $^1(n_O\pi^*)$ and $^3(n_O\pi^*)$ states are usually very small (below 10 cm$^{-1}$). However, at the three-state $S_1/T_2/T_1$ near-degeneracy $T_2$ acquires a minor contribution of $\pi\pi^*$ character coming from the $T_1$ state, enhancing the SOCs to 30 cm$^{-1}$ on average and a maximum of more than 40 cm$^{-1}$. Such strong SOCs together with the small energy differences between $S_1$ and $T_2$ lead to a significant singlet-triplet mixing, so that the resulting states are neither pure singlets (total spin expectation value $\langle\hat{S}^2\rangle$=0.0) nor pure triplets ($\langle\hat{S}^2\rangle$=2.0). For the \emph{keto} tautomer, which shows strong mixing, the values of $\langle\hat{S}^2\rangle$ are in the range 0.1 and 1.9 (in atomic units) for approximately 7\% of the simulation time.
The finding that the $^1(n\pi^*)$ state is a precursor to triplet formation was already suggested by Hare et al.,\cite{hare_internal_2007} although their work focused on excited-state dynamics in solution.
% ISC takes place between the $S_1$ and $T_2$ because although at the points of the crossing both states are mainly of $n_O\pi^*$ character, the $T_2$ state gets a minor contribution of $\pi\pi^*$ character coming from the $T_1$ state, enhancing the SOCs.
% The combination of large SOCs and the fact that the $n_O\pi^*$ singlet and triplet states are extremely close in energy (usually on the order of 0.01 eV) for extended regions of the $S_1$ PES enables efficient ISC. Moreover, because ISC to $T_2$ requires the proximity of the $T_1$ state,  $T_2\rightarrow T_1$ IC is very much favored at these geometries, as already observed in Ref.~\cite{richter_femtosecond_2012}.

The analysis of all the geometries at which ISC takes place also reveals that an elongated \ce{C2=O} bond and a short \ce{C2-N3} bond seems to be the key feature of the singlet-triplet crossings, while pyramidalization of the amino group -- albeit present -- is not important. A pyramidalization of the \ce{N1} atom may also be relevant.
Richter et al.\cite{richter_femtosecond_2012} also stated that pyramidalization at the amino group and at the \ce{N1} atom is a key element in the ISC process. However, their analysis is based on all geometries which exhibited a very small singlet-triplet gap, regardless whether an actual hop occurred. In the current work, all singlet-triplet hops were analyzed case-by-case, giving a much more detailed picture of the ISC path. This examination showed that amino group pyramidalization does occur, but is probably coincidental and not responsible for the interaction of singlet and triplet states. Instead, we find that all geometries where ISC occurred show an elongated \ce{C2=O} bond and a short \ce{C2-N3} bond.
Interestingly, the hopping geometries possess \ce{C2=O} and \ce{C2-N3} bond lengths similar to the \emph{semi-planar} $S_1/S_0$ and $S_2/S_1/S_0$ CoIns.
Since in the case of \emph{keto} cytosine ISC necessitates the near-degeneracy of three states ($S_1$, $T_1$ and $T_2$), optimization of either $S_1/T_1$ or $S_1/T_2$ crossings did not lead to geometries representative of the ISC mechanism. Nevertheless, we successfully optimized a $T_2/T_1$ crossing, see Fig.~\ref{fig:geom_keto}f. At this geometry, $S_1$ is only 0.05 eV above $T_1$ and $T_2$ and SOCs are large (26 and 37cm$^{-1}$), as discussed above. This geometry is characterized by a pyramidalization at the \ce{N1} atom, and it is therefore labelled \emph{\ce{N1}-pyramidalized} CoIn.

%\textbf{Consequently, ISC can lead either directly to $T_1$ or to $T_2$ followed by immediate triplet IC.}

As previously found by Richter et al.,\cite{richter_femtosecond_2012} the ISC pathway discussed above differs from the one predicted by quantum chemistry by Merch\'an et al.\cite{merchan_triplet-state_2005}, which was proposed to be $S_1\rightarrow T_1$. In the same paper, it is already discussed that the SOC enhancement is due to a $n\pi^*$/$\pi\pi^*$ mixing. They proposed that this mixing is induced by the pyramidalization of \ce{C6}. Even though we find that SOC is indeed enhanced by such a mixing, we cannot confirm the importance of the \ce{C6} pyramidalization. Here instead, the singlet-triplet crossing are mediated by the elongation of the \ce{C2=O} bond and a shortening of the \ce{C2-N3} bond, as discussed above.
The same authors~\cite{merchan_triplet-state_2005} calculate an ISC probability $P_{\text{ISC}}$ as 0.1\% per passage of the singlet-triplet crossing region along their pathway by means of a Landau-Zener type model. In this model,\cite{riad_manaa_mechanism_1991} $P_{\text{ISC}}$ is given by:%
\begin{align}
  P_{\text{ISC}}&=1-\text{e}^{-\frac{\pi}{4}\xi},\label{eq:LZ1}\\
  \xi&=\frac{8}{\hbar\mathbf{g}\cdot\mathbf{v}}\left|\left\langle\Psi_1\middle|\hat{H}_{SO}\middle|\Psi_2\right\rangle\right|^2,\label{eq:LZ2}
\end{align}
where $\mathbf{g}$ is the gradient difference vector, $\mathbf{v}$ is the velocity vector of the nuclei and their scalar product is the change in energy difference with respect to time $\Delta\Delta E/\Delta t$.
Based on energetic arguments,\cite{merchan_triplet-state_2005} Merch\'an et al. estimate $\Delta\Delta E/\Delta t$=0.1 eV/fs, which assumes that the system is moving perpendicular to the singlet-triplet seam. Our simulations reveal that the system is instead moving along the singlet-triplet seam for an extended time; accordingly, we expect $\Delta\Delta E/\Delta t$ to be much smaller. In such a case, the ISC probability corresponding to a single passage of the singlet-triplet crossing region would result well above 1\%.
Over the course of many such passages and combined with the apparent irreversibility of ISC (we do not observe $T\rightarrow S$ transitions) population steadily accumulates in the triplet states.

% \begin{itemize}
%   \item sub-ps ISC, scale of a few 100 fs, disagrees with martin, who observed tens of fs ISC
%   \item reason: much more stable propagation wrt SOC
%   \item since in this work $S_2\rightarrow S_1$ is faster than ISC (because ISC is not as fast as before), there is no $S_2$ to $T$
%   \item richter stated that NH2 group and N1H pyramidalization are responsible for ISC, and observation is only based on energy gaps, not actual hops
%   \item now we performed an trajectory-by-trajectory analysis of the ISC mechanism
%   \item still find NH2 and N1H, however, they seem to be coincidental and not responsible for the interaction of S and T states
%   \item instead we see a clear dependence on the NC=O group
%
%   \item nevertheless, we agree in that the mechanism is basically $S_1\rightarrow T_2\rightarrow T_1$, where $S_1$ and $T_2$ are npi
%   \item large SOC emerge if other states than these two come close and wavefunction character mixes => more
%   \item after the ISC process, we observe a very fast decay to $T_1$ (agreeing with richter), since it is already close when the ISC hop occurs
%
%   \item in agreement with merchan CO bond is stretched, pyramidalization is not important
%   \item but otherwise is different, t2 is very important in ISC procss
%
%   \item LZ stuff
% \end{itemize}
% =========================================================================================================================================== %

\subsection{Relaxation mechanism of enol cytosine}\label{ssec:enol}

\begin{figure}
  \centering
  \includegraphics[scale=1]{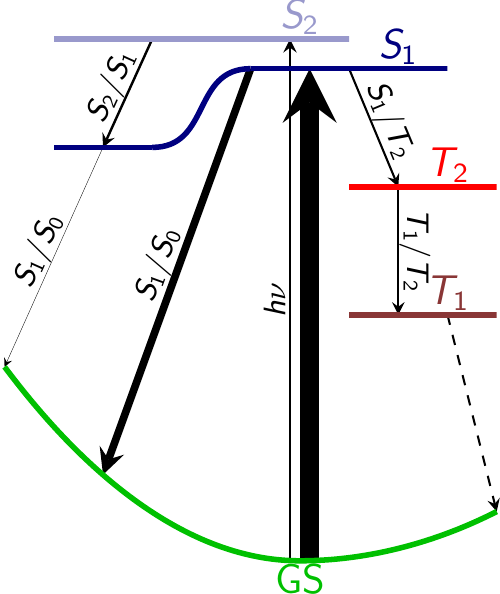}
  \caption{Schematic representation of the relaxation pathways observed in the \emph{enol} tautomer after excitation (indicated by $h\nu$). Arrow thickness depicts the fraction of trajectories taking a particular pathway. ISC from $T_1$ to $S_0$ is not observed in the present work, only assumed; this is indicated by a dashed line.}
  \label{fig:overview_enol}
\end{figure}

As already discussed above  \emph{enol} cytosine shows a dramatically different dynamics than \emph{keto} cytosine.  Figure~\ref{fig:overview_enol} collects the observed relaxation pathways, revealing a less complicated behaviour compared to the \emph{keto} tautomer.
To the best of our knowledge, there are no CoIns reported for the \emph{enol} tautomer. Therefore, special attention was put to analyze all relaxation pathways and optimize the related CoIns and singlet-triplet crossings (at the SA3S+4T-CASSCF(12,9)/6-31G* level of theory).

\begin{figure}
  \centering
  \includegraphics[scale=1]{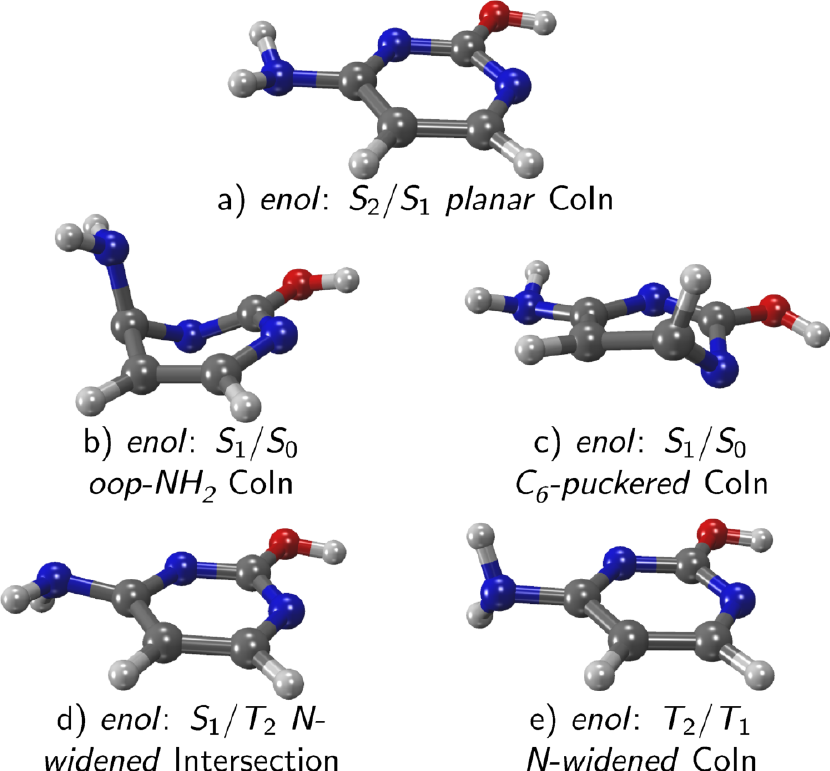}
  \caption{Optimized (conical) intersections of the \emph{enol} tautomer at the SA3S+4T-CASSCF(12,9)/6-31G* level of theory. The coordinates of all geometries are given in the Supporting Information.}
  \label{fig:geom_enol}
\end{figure}

\begin{figure}
  \centering
  \includegraphics[scale=1]{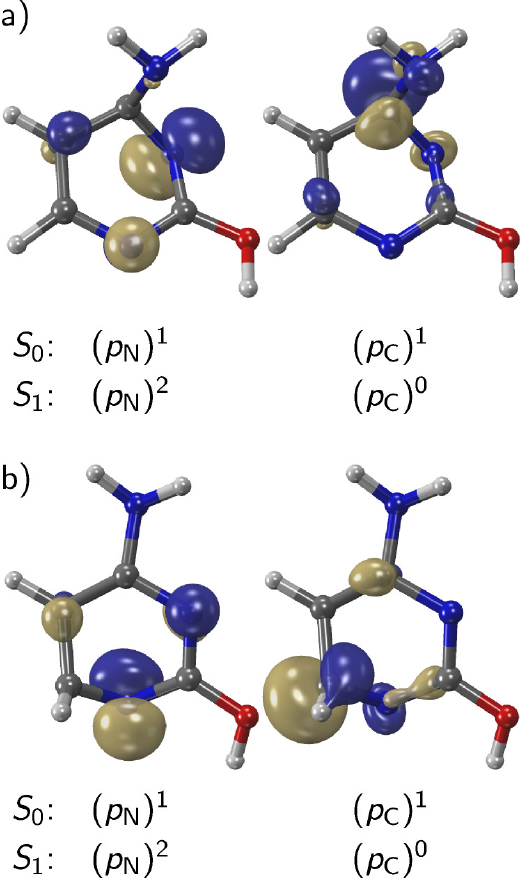}
  \caption{Orbitals for the conical intersection a) \emph{oop-\ce{NH2}} and b) \emph{\ce{C6}-puckered} in \emph{enol} cytosine.}
  \label{fig:orb_enol}
\end{figure}

The deactivation of the $S_2$ population proceeds via a cascade of CoIn, first going to the $S_1$ and from the $S_1$ to the $S_0$. The geometry of the CoIn responsible for $S_2\rightarrow S_1$ interconversion (Fig.~\ref{fig:geom_enol}a) very much resembles the ground state minimum and it is labelled $S_2/S_1$ \emph{planar} CoIn.  The two involved electronic states are of $\pi\pi^*$ ($S_2$) and $n\pi^*$ ($S_1$) character, thus the $S_2\rightarrow S_1$ IC is part of the $\pi\pi^*\rightarrow n\pi^*$ transition discussed in section~\ref{ssec:lifetimes}.
Two CoIns have been identified as responsible for the relaxation from the $S_1$ state to the ground state. The first one (Fig. \ref{fig:geom_enol}b) is termed \emph{oop-\ce{NH2}} CoIn, in analogy to the CoIn of \emph{keto} cytosine. It is characterized by a semi-twist geometry with puckering of atoms \ce{N3} and \ce{C4}, as well as a strong out-of-plane deformation of the amino-group.% (\emph{oop-\ce{NH2}}).
The relevant orbitals for the $S_0$ and $S_1$ states are given in Fig. \ref{fig:orb_enol}a. This CoIn is similar to the one of ethylene, where the twist around the double bond and pyramidalization at one \ce{C} atom leads to the crossing of ground and excited state. The minimum of the intersection seam was optimized at 4.34 eV, which is much lower than the excitation energy, making the CoIn accessible.
The second CoIn (Fig. \ref{fig:geom_enol}c) is the equivalent of the \emph{\ce{C6}-puckered} CoIn in the \emph{keto} tautomer, with puckering of the atoms \ce{N1} and \ce{C6}. The orbitals are given in Fig. \ref{fig:orb_enol}b, showing that this CoIn is also equivalent to an ethylenic CoIn, but here the twist does not involve the C-C bond but the C-N one. Even though this CoIn is higher in energy than the \emph{oop-\ce{NH2}} CoIn (4.64 vs. 4.34 eV), it accounts for the majority of relaxing trajectories (70\% of the trajectories reaching the ground state within 1 ps). One reason might be the fact that \ce{N1} and \ce{C6} carry very light or no side groups, giving the \emph{twist-\ce{N1C6}} normal mode a lower reduced mass and a higher oscillation period. Thus, the \emph{\ce{C6}-puckered} CoIn can be approached more often and this relaxation pathway becomes more important.

Compared to the \emph{keto} tautomer, ISC is much slower in the \emph{enol} form. This is because there are no extended areas on the PES where singlets and triplets are close to each other. Additionally, the \emph{enol} form lacks the carbonyl group which is responsible for the large SOCs in the \emph{keto} tautomer. As a consequence, the average SOCs are smaller by a factor of more than two in the \emph{enol} form and $S$/$T$ mixing is much less pronounced, with $\langle S^2\rangle$ between 0.1 and 1.9 (in atomic units) only for 1\% of the simulation time (compare to 7\% for the \emph{keto}).
Yet the $S_1\rightarrow T_2$ ISC pathway contributes to a minor extent. The geometry of the $S_1/T_2$ \emph{\ce{N}-widened} crossing (Fig.~\ref{fig:geom_enol}d) is planar and shows comparably large \ce{C6-N1-C2} and \ce{C2-N3-C4} angles. The interacting states $S_1$ and $T_2$ are of $\pi\pi^*$ and $n\pi^*$ character, respectively. The energy of this crossing was found to be 4.79 eV and SOCs are close to 10cm$^{-1}$. In all the simulations, this ISC was followed by triplet IC ($T_2\rightarrow T_1$) within 40 fs on average, facilitating a change of wavefunction character from $n\pi^*$ to $\pi\pi^*$. The relevant $T_2/T_1$ CoIn, depicted in Fig.~\ref{fig:geom_enol}e, shows a similar ring structure as the $S_1/T_2$ crossing geometry (Fig.~\ref{fig:geom_enol}d), albeit with slightly smaller angles at the nitrogen atoms. The $T_2\rightarrow T_1$ CoIn in the \emph{enol} tautomer was predicted at 4.40 eV.

% =========================================================================================================================================== %
% =========================================================================================================================================== %
% =========================================================================================================================================== %

\section{Conclusion}\label{sec:conclusion}

We present results of ab initio surface-hopping dynamics including singlet and triplet states to unravel the relaxation mechanism of the \emph{keto} and \emph{enol} tautomers of cytosine after light irradiation. The simulations show an approximately biexponential decay with time constants of 6 and 230 fs for the \emph{keto} tautomer and a monoexponential decay with a time constant of 2400 fs for the \emph{enol} tautomer.

It is proposed in this work that the $T_1$ state of both tautomers lie too low in energy to be detected by usual probe laser setups (e.g. 3$\times$800 nm ionization). Thus, the experimentally measured decay rates arise from the simultaneous decay to the $S_0$ ground state and ISC followed by IC to $T_1$. Under this premise, for the \emph{keto} tautomer the two decay constants are 7 and 270 fs, where the second is an effective time constant for both processes. The relaxation of the \emph{enol} tautomer is still monoexponential with a time constant of 1900 fs. Additionally, for the \emph{enol} form it is found that the $S_1$ state changes with a time constant of 40 fs from bright to dark, which is due to an adiabatic change of wavefunction character from $\pi\pi^*$ to $n\pi^*$. The calculated decay times (7, 270 and 1900 fs) agree well with the available experimental results. It is therefore proposed that while the \emph{keto} tautomer is responsible for the reported ultrafast transients, the \emph{enol} tautomer could contribute to the ps time scale measured experimentally.

For the \emph{keto} tautomer, a number of IC and ISC competing processes are found. The relaxation mechanism involving triplet states was found to be $S_1\rightarrow T_2\rightarrow T_1$. Even though both $S_1$ and $T_2$ states are predominantly of $n_O\pi^*$ character, SOC between the two states can be dramatically increased if additional states are close allowing for mixing with $\pi\pi^*$ character. Additionally, the small energy differences between $S_1$ and $T_2$ of less than 0.01 eV allow for effective ISC to take place. One unambiguously identified structural feature promoting ISC is the stretching of the \ce{C=O} group.

The relaxation mechanism of \emph{enol} cytosine is considerably simpler than the \emph{keto} counterpart and ground state repopulation is significantly slower than in \emph{keto} cytosine. Even though the \emph{enol} form exhibits the \emph{\ce{C6}-puckered} CoIn, it lacks the carbonyl group and thus the very efficient pathway associated with the stretching of the \ce{C=O} bond. Also due to the missing carbonyl group, there is no efficient ISC channel found in \emph{enol} cytosine.

% =========================================================================================================================================== %
% =========================================================================================================================================== %
% =========================================================================================================================================== %

\section*{Acknowledgements}\label{sec:acknowledgments}

This work is supported by the Deutsche Forschungsgemeinschaft (DFG) within the Project GO 1059/6-1 and by the German Federal Ministry of Education and Research within the Research Initiative PhoNa. Generous allocation of computer time at the Vienna Scientific Cluster (VSC) is gratefully acknowledged. The authors wish to thank Tom Weinacht for useful discussions.

\section*{Supporting information}\label{sec:SI}
\section{Optimized Geometries}

\subsection{Keto tautomer}

Ground state $S_0$ minimum from MP2/6-311G**:
\begin{center}
  \linespread{1}
  \begin{verbatim}
13
keto cytosine, S0 minimum
C -0.068773583  2.644406855 -0.017986395
C -1.217618868  1.778517632 -0.069019606
C -0.954950130  0.443464687 -0.074742685
N  0.330827750  0.015058300 -0.026586955
C  1.457239615  0.878304672  0.028936022
N  1.177876864  2.231412989  0.022285695
O  2.570795443  0.392123670  0.068829626
N -0.276445039  3.999250355  0.054355190
H -1.124704501  4.356775579 -0.356071716
H  0.550514655 -0.971271203 -0.029503369
H -2.230607364  2.158058927 -0.090139977
H -1.726612162 -0.318669951 -0.112825632
H  0.555433319  4.545287487 -0.120546198
  \end{verbatim}
\end{center}

Triplet $T_1$ minimum from SA4S+3T-CASSCF(12,9)/6-31G*:
\begin{center}
  \linespread{1}
  \begin{verbatim}
13
keto cytosine, T1 minimum
C -0.155431664  2.663756914 -0.060533707
C -1.217459954  1.785980393 -0.009540096
C -0.976302678  0.322212442  0.008572157
N  0.373013888 -0.009261542  0.063766077
C  1.400362047  0.889012469 -0.072594365
N  1.130514120  2.255109708 -0.111011737
O  2.548185488  0.508610768 -0.134948306
N -0.339914463  4.043258092 -0.152964581
H -1.107538117  4.394659490  0.378046722
H  0.648739212 -0.965379403  0.074841606
H -2.227609632  2.145273300 -0.042475023
H -1.630095777 -0.310222691  0.582978603
H  0.499252796  4.546469965  0.043969268
  \end{verbatim}
\end{center}

$S_1/S_0$ conical intersection (equivalent to \emph{semi-planar}\cite{barbatti_photodynamical_2011} or \emph{$R_x$(ci01)'}\cite{kistler_radiationless_2007}, also found in Refs. \onlinecite{blancafort_energetics_2007,hudock_excited-state_2008,merchan_ultrafast_2003,blancafort_singlet_2005,ismail_ultrafast_2002}) from SA4S+3T-CASSCF(12,9)/6-31G* ( displayed in the paper in Figure~8 b)):
\begin{center}
  \linespread{1}
  \begin{verbatim}
13
keto cytosine, S1/S0 CoIn semi-planar
C -0.166396960  2.640760894 -0.079394749
C -1.192810242  1.779684135 -0.173074550
C -0.951578474  0.337422016 -0.133105868
N  0.398924317 -0.012994102  0.058264395
C  1.324761162  0.965285998 -0.001526212
N  1.214951868  2.182753340 -0.162727951
O  2.531287750  0.261401662 -0.022012719
N -0.226595732  4.008665477  0.174147530
H -1.133114629  4.403857555  0.035658729
H  0.633042609 -0.841855865  0.559662143
H -2.210994922  2.123427874 -0.146118115
H -1.555836958 -0.369128226 -0.669576111
H  0.465534794  4.512510469 -0.341611237
  \end{verbatim}
\end{center}

$S_1/S_0$ conical intersection (equivalent to \emph{oop-\ce{NH2}}\cite{barbatti_photodynamical_2011} or \emph{$R_x$(ci01)$_{\text{sofa}}$}\cite{kistler_radiationless_2007}) from SA4S+3T-CASSCF(12,9)/6-31G* (displayed in the paper in Figure~8 c)):
\begin{center}
  \linespread{1}
  \begin{verbatim}
13
keto cytosine, S1/S0 CoIn oop-NH2
C -0.203970465  2.727356856 -0.383211967
C -1.325375587  1.801421910 -0.144453849
C -0.969223920  0.549275706  0.220007610
N  0.378205602  0.160706084  0.284467185
C  1.394703115  1.009498485 -0.118412326
N  0.870812295  2.021274691 -0.942208223
O  2.551907818  0.828359921  0.106215686
N  0.137804078  3.637156589  0.633581946
H  0.750785095  4.361500435  0.317502996
H  0.646017785 -0.560401498  0.916486860
H -2.350495248  2.116971290 -0.150898418
H -1.676090982 -0.216689427  0.474761683
H -0.661639361  4.038674752  1.079074339
  \end{verbatim}
\end{center}

$S_2/S_1/S_0$ conical intersection (equivalent to \emph{$R_{x2}$(ci012)}\cite{kistler_three-state_2008}, also found in Refs. \onlinecite{blancafort_key_2004,gonzalez-vazquez_time-dependent_2010}) from SA4S+3T-CASSCF(12,9)/6-31G* (optimized as $S_2/S_0$ CoIn, displayed in the paper in Figure~8 a)):
\begin{center}
  \linespread{1}
  \begin{verbatim}
13
keto cytosine, S2/S1/S0 CoIn
C -0.261439224  2.664725400  0.054200961
C -1.265412911  1.799284410  0.080040150
C -0.956569139  0.334325871  0.177939405
N  0.372862542  0.063676246 -0.006735474
C  1.282220844  1.026515156  0.010256654
N  1.141836817  2.266708140  0.040109599
O  2.525958770  0.337847195 -0.041301157
N -0.344624335  4.041831255  0.089946437
H -1.218022027  4.430105730 -0.192364480
H  0.707582112 -0.877872174 -0.019819596
H -2.288473732  2.118574384  0.133300383
H -1.637603083 -0.426674264 -0.149825303
H  0.436617774  4.483078555 -0.346405359
  \end{verbatim}
\end{center}

$S_2/S_1$ conical intersection (equivalent to \emph{$R_{x}$(ci12)}\cite{kistler_radiationless_2007}) from SA4S+3T-CASSCF(12,9)/6-31G* (displayed in the paper in Figure~8 d)):
\begin{center}
  \linespread{1}
  \begin{verbatim}
13
keto cytosine, S2/S1 CoIn Rx(ci12)
C -0.149370274  2.678410912  0.015783698
C -1.205967219  1.831835630 -0.080894267
C -0.964238436  0.416235123 -0.113392905
N  0.356222646  0.035935168 -0.056733508
C  1.343226236  0.955580018  0.034615799
N  1.190981536  2.222580917  0.074306490
O  2.543743488  0.361351984  0.078599886
N -0.237921995  4.060335838  0.129872648
H -1.062845499  4.460854378 -0.262232201
H  0.626336764 -0.921633594 -0.075489298
H -2.215274002  2.194774217 -0.116693675
H -1.708546500 -0.344936807 -0.183183716
H  0.587698934  4.517013432 -0.195492554
  \end{verbatim}
\end{center}

$S_2/S_1$ conical intersection (equivalent to \emph{$R_{x}$(ci12)'}\cite{kistler_radiationless_2007}) from SA4S+3T-CASSCF(12,9)/6-31G* (displayed in the paper in Figure~8 e)):
\begin{center}
  \linespread{1}
  \begin{verbatim}
13
keto cytosine, S2/S1 CoIn Rx(ci12)'
C -0.187608306  2.693660927 -0.034300984
C -1.233389808  1.820140729 -0.065640134
C -0.974744749  0.421394812 -0.059158764
N  0.341022247 -0.011546732 -0.127368846
C  1.409876157  0.857221764  0.023840874
N  1.102522343  2.147911136  0.052207979
O  2.548444403  0.400405770  0.117565055
N -0.235655190  4.081893960  0.049556768
H -1.122571101  4.456578018 -0.213060947
H  0.564099281 -0.976966949 -0.032296910
H -2.246086930  2.174692340 -0.090583877
H -1.735842196 -0.327702079 -0.037524089
H  0.496372989  4.521795555 -0.470340399
  \end{verbatim}
\end{center}

$T_2/T_1$ conical intersection from SA4S+3T-CASSCF(12,9)/6-31G* (displayed in the paper in Figure~8 f)):
\begin{center}
  \linespread{1}
  \begin{verbatim}
13
keto cytosine, T2/T1 CoIn N-pyramidalized
C -0.089596812  2.668085713  0.050209979
C -1.179344329  1.768997116 -0.040720322
C -0.942732156  0.427634092 -0.119732290
N  0.393733641 -0.086690561 -0.039519900
C  1.341083046  0.936068683  0.023448638
N  1.194638688  2.178587886  0.024328469
O  2.606879229  0.476202182  0.244287131
N -0.224401703  4.042977997  0.078907222
H  0.589361321  4.605753197  0.137930080
H  0.588253733 -0.754201858 -0.757671993
H -2.190562081  2.129928302 -0.033322784
H -1.709414850 -0.317876679 -0.159502431
H -1.086567565  4.479574366 -0.134406515
  \end{verbatim}
\end{center}

\subsection{Enol tautomer}

Ground state $S_0$ minimum from MP2/6-311G**:
\begin{center}
  \linespread{1}
  \begin{verbatim}
13
enol cytosine, S0 minimum
C  0.010625510  1.766522530  0.006612200
C -2.351501000  0.536787460 -0.009538460
C -2.294608210 -2.077209830 -0.011090920
C  1.965306570 -2.029622620 -0.004959960
N  2.185931430  0.479640730  0.013035510
N -0.136759500 -3.421963010 -0.012856160
N  0.211629730  4.355569160 -0.107217780
O  4.173197140 -3.292874090 -0.002123440
H  1.940634190  5.001467260  0.380212980
H -1.244475120  5.349830620  0.615080210
H -4.116243340  1.576035790 -0.028768870
H -4.036258760 -3.168440630 -0.015561430
H  3.686570600 -5.050999310 -0.008886730
  \end{verbatim}
\end{center}

Triplet $T_1$ minimum from SA3S+4T-CASSCF(12,9)/6-31G*:
\begin{center}
  \linespread{1}
  \begin{verbatim}
13
enol cytosine, T1 minimum
C  0.075811830  0.978202089  0.074726012
C -1.233845434  0.320292344  0.238339781
C -1.262467895 -1.096970361 -0.117484128
C  0.998015466 -1.109436149  0.192452434
N  1.150510272  0.285061248  0.007920082
N -0.183158735 -1.796628489 -0.184064280
N  0.089943272  2.344567037 -0.127690583
O  2.104094655 -1.816422755 -0.102489975
H  1.005543173  2.740161530 -0.086794785
H -0.573972377  2.864221410  0.405452301
H -2.136005944  0.892368209  0.327075565
H -2.196083640 -1.570959723 -0.360977993
H  2.848420810 -1.229311101 -0.064943926
  \end{verbatim}
\end{center}

$S_1/S_0$ conical intersection (equivalent to \emph{oop-\ce{NH2}}\cite{barbatti_photodynamical_2011} or \emph{$R_x$(ci01)$_{\text{sofa}}$}\cite{kistler_radiationless_2007}) from SA3S+4T-CASSCF(12,9)/6-31G* (displayed in the paper in Figure~11 b)):
\begin{center}
  \linespread{1}
  \begin{verbatim}
13
enol cytosine, S1/S0 CoIn oop-NH2
C -0.038436380  0.980182354 -0.434852906
C -1.350283143  0.306708928 -0.224081598
C -1.239981620 -0.954355639  0.257346915
C  1.032346993 -1.026127374 -0.080495536
N  0.858178803  0.028641156 -0.948198860
N  0.005449774 -1.598417717  0.426929157
N  0.454502002  1.687084920  0.682530738
O  2.249367213 -1.557436180 -0.017644652
H -0.246493322  2.233488430  1.139539029
H  1.239013912  2.265850689  0.459166660
H -2.288385862  0.825666585 -0.275998791
H -2.089261540 -1.544003481  0.542393979
H  2.186902596 -2.377930095  0.459800540
  \end{verbatim}
\end{center}

$S_1/S_0$ conical intersection (equivalent to \emph{\ce{C6}-puckered}\cite{barbatti_photodynamical_2011} or \emph{$R_x$(ci01)$_{\text{twist}}$}\cite{kistler_radiationless_2007}) from SA3S+4T-CASSCF(12,9)/6-31G* (displayed in the paper in Figure~11 c)):
\begin{center}
  \linespread{1}
  \begin{verbatim}
13
enol cytosine, S1/S0 CoIn C6-puckered
C -0.080344470  0.887657845 -0.017436774
C -1.289403885  0.263338890 -0.052368262
C -1.159845161 -1.179485616  0.233822828
C  1.111240830 -1.081875281 -0.011282940
N  1.145078352  0.184443933  0.141650860
N -0.064441447 -1.627150810 -0.493821746
N  0.121715425  2.251762149 -0.148699921
O  2.222215172 -1.811343424  0.039880656
H  0.954291743  2.571355938  0.297525897
H -0.669802671  2.820931925  0.060909518
H -2.217839364  0.800585928 -0.050519164
H -1.218341526 -1.493135739  1.272105888
H  2.030917671 -2.707896804 -0.206212829
  \end{verbatim}
\end{center}

$S_2/S_1$ conical intersection from SA3S+4T-CASSCF(12,9)/6-31G* (displayed in the paper in Figure~11 a)):
\begin{center}
  \linespread{1}
  \begin{verbatim}
13
enol cytosine, S2/S1 CoIn planar
C -0.026790111  0.946653018 -0.010610954
C -1.271718210  0.295706614 -0.002721522
C -1.279297066 -1.136601501 -0.024759159
C  1.070143536 -1.086769538 -0.051369587
N  1.154663446  0.238841708 -0.051463563
N -0.063837079 -1.787716919 -0.035686683
N  0.112087082  2.321372596 -0.064265465
O  2.220822122 -1.755515465 -0.068726004
H  1.012783253  2.639467537  0.224416193
H -0.627078841  2.843440771  0.354401580
H -2.187735625  0.854761206  0.005452497
H -2.162380236 -1.735471503  0.027833850
H  2.027839514 -2.685391116 -0.075515400
  \end{verbatim}
\end{center}

$S_1/T_2$ intersection from SA3S+4T-CASSCF(12,9)/6-31G* (displayed in the paper in Figure~11 a)):
\begin{center}
  \linespread{1}
  \begin{verbatim}
13
enol cytosine, S1/T2 In N-widened
C -0.058318459  0.962672094 -0.046222626
C -1.263764473  0.295666889 -0.100567801
C -1.256172437 -1.103341970 -0.018251284
C  1.116306760 -1.148474943  0.091613178
N  1.075157953  0.153462630  0.035445844
N -0.022240460 -1.752015986  0.040877979
N  0.184114678  2.342056939  0.001196759
O  2.279543593 -1.773657058  0.183316676
H -0.663020101  2.862446079  0.023757858
H  0.735676195  2.657308754 -0.769432194
H -2.186586241  0.842515491 -0.180636639
H -2.124276021 -1.721355705 -0.030588070
H  2.112437670 -2.707810984  0.209922950
  \end{verbatim}
\end{center}

$T_2/T_1$ conical intersection from SA3S+4T-CASSCF(12,9)/6-31G* (displayed in the paper in Figure~11 e)):
\begin{center}
  \linespread{1}
  \begin{verbatim}
13
enol cytosine, T2/T1 CoIn N-widened
C -0.014436349  0.984136324 -0.027760212
C -1.190329903  0.363885161  0.355802568
C -1.206674696 -1.023842817  0.512020065
C  1.054764580 -1.183508330 -0.071290605
N  1.095810940  0.137052237 -0.234598094
N -0.038928880 -1.765153007  0.288854810
N  0.131426673  2.367729979 -0.227957888
O  2.176266760 -1.858047984 -0.295481979
H  0.394111518  2.580785566 -1.171593384
H  0.826303176  2.759032951  0.379074486
H -2.065997294  0.960056132  0.524952285
H -2.070141781 -1.582254685  0.802396469
H  2.001762739 -2.780481702 -0.147990232
  \end{verbatim}
\end{center}

% 
% In the supporting information, the optimized geometries for different PES minima, CoIns and singlet-triplet intersections can be found in xyz format.

% =========================================================================================================================================== %
% =========================================================================================================================================== %
% =========================================================================================================================================== %

%%%% This bibstyle tries to emulate the CHEMPHYSCHEM citation style %
%\bibliographystyle{chemphyschem}                                    %
%%%% -------------------------------------------------------------- %

%\bibliography{cytosine_paper}

% =========================================================================================================================================== %
% =========================================================================================================================================== %
% ===========================================================================================================================================

\end{document}